%% file: aanda.tex
\documentclass[longauth]{aa}
\usepackage{graphicx}
\usepackage{txfonts}

\usepackage[toc,page]{appendix}

\usepackage[table]{xcolor}
\usepackage[utf8]{inputenc} 
\usepackage[T1]{fontenc}    
\usepackage{hyperref}       
\usepackage{url}            
\usepackage{booktabs}       
\usepackage{amsfonts}       
\usepackage{nicefrac}       
\usepackage{microtype}      
\usepackage{lipsum}		
\usepackage{footmisc}
\usepackage{graphicx}
\usepackage{natbib}
\usepackage{tabu}
\usepackage{makecell}
\usepackage{listings}
\usepackage{amsmath}
\usepackage{color}
\usepackage{multirow}
\usepackage{xcolor}
\usepackage{caption}
\usepackage[font={small},labelfont={bf}]{caption}

\usepackage{float}
\floatstyle{plaintop}
\restylefloat{table}

\newcommand{\angstrom}{\text{\normalfont\AA}}
\usepackage{euclid}
\usepackage{placeins}
\usepackage{cleveref}

\usepackage{ulem}

\newcommand*{\spr}{{\tt SPRITZ}\,}

\definecolor{anti-flashwhite}{rgb}{0.95, 0.95, 0.96}
\definecolor{hotpink}{rgb}{1.0, 0.41, 0.71}
\definecolor{fuchsia}{rgb}{1.0, 0.0, 1.0}
\definecolor{violet(ryb)}{rgb}{0.53, 0.0, 0.69}

\defcitealias{Weaver2022}{W22}
    \makeatletter
\def\@fnsymbol#1{\ensuremath{\ifcase#1\or *\or \dagger\or \ddagger\or
   \mathsection\or \mathparagraph\or \|\or **\or \dagger\dagger
   \or \ddagger\ddagger \else\@ctrerr\fi}}
    \makeatother

\begin{document}

\title{\Euclid: Identifying the reddest high-redshift galaxies in the Euclid
Deep Fields with gradient-boosted trees\thanks{This paper is published on
behalf of the Euclid Consortium.}}

\authorrunning{T. Signor et al.}

\titlerunning{Identifying the reddest high-$z$ galaxies in the Euclid Deep Fields with gradient-boosted trees}
\newcommand{\orcid}[1]{} 

\author{T.~Signor\orcid{0009-0003-5121-3567}$^{1,2,3}$, G.~Rodighiero\orcid{0000-0002-9415-2296}$^{3,4}$\thanks{\email{giulia.rodighiero@unipd.it}}, L.~Bisigello\orcid{0000-0003-0492-4924}$^{3,4}$, M.~Bolzonella\orcid{0000-0003-3278-4607}$^{5}$, K.~I.~Caputi$^{6,7}$, E.~Daddi\orcid{0000-0002-3331-9590}$^{8}$, G.~De~Lucia\orcid{0000-0002-6220-9104}$^{9}$, A.~Enia\orcid{0000-0002-0200-2857}$^{10,11}$, L.~Gabarra$^{12}$, C.~Gruppioni\orcid{0000-0002-5836-4056}$^{5}$
% ...
}

\institute{$^{1}$ Instituto de Estudios Astrof\'isicos, Facultad de Ingenier\'ia y Ciencias, Universidad Diego Portales, Av. Ejercito 441, Santiago, Chile\\
$^{2}$ Inria Chile Research Center, Av. Apoquindo 2827, piso 12, Las Condes, Santiago, Chile\\
$^{3}$ Dipartimento di Fisica e Astronomia "G. Galilei", Universit\`a di Padova, Via Marzolo 8, 35131 Padova, Italy\\
$^{4}$ INAF-Osservatorio Astronomico di Padova, Via dell'Osservatorio 5, 35122 Padova, Italy\\
$^{5}$ INAF-Osservatorio di Astrofisica e Scienza dello Spazio di Bologna, Via Piero Gobetti 93/3, 40129 Bologna, Italy\\
$^{6}$ Kapteyn Astronomical Institute, University of Groningen, PO Box 800, 9700 AV Groningen, The Netherlands\\
$^{7}$ Cosmic Dawn Center (DAWN)\\
$^{8}$ AIM, CEA, CNRS, Universit\'{e} Paris-Saclay, Universit\'{e} de Paris, 91191 Gif-sur-Yvette, France\\
$^{9}$ INAF-Osservatorio Astronomico di Trieste, Via G. B. Tiepolo 11, 34143 Trieste, Italy\\
$^{10}$ Dipartimento di Fisica e Astronomia, Universit\`a di Bologna, Via Gobetti 93/2, 40129 Bologna, Italy\\
$^{11}$ INAF-IASF Bologna, Via Piero Gobetti 101, 40129 Bologna, Italy\\
$^{12}$ Department of Physics, Oxford University, Keble Road, Oxford OX1 3RH, UK\\
$^{13}$ Instituto de Astrof\'isica e Ci\^encias do Espa\c{c}o, Universidade do Porto, CAUP, Rua das Estrelas, PT4150-762 Porto, Portugal\\
$^{14}$ DTx -- Digital Transformation CoLAB, Building 1, Azur\'em Campus, University of Minho, 4800-058 Guimar\~aes, Portugal\\
$^{15}$ Department of Mathematics and Physics, Roma Tre University, Via della Vasca Navale 84, 00146 Rome, Italy\\
$^{16}$ INAF-IASF Milano, Via Alfonso Corti 12, 20133 Milano, Italy\\
$^{17}$ School of Physical Sciences, The Open University, Milton Keynes, MK7 6AA, UK\\
$^{18}$ INAF-Istituto di Astrofisica e Planetologia Spaziali, via del Fosso del Cavaliere, 100, 00100 Roma, Italy\\
$^{19}$ INAF-Osservatorio Astronomico di Brera, Via Brera 28, 20122 Milano, Italy\\
$^{20}$ INFN-Sezione di Bologna, Viale Berti Pichat 6/2, 40127 Bologna, Italy\\
$^{21}$ Max Planck Institute for Extraterrestrial Physics, Giessenbachstr. 1, 85748 Garching, Germany\\
$^{22}$ Universit\"ats-Sternwarte M\"unchen, Fakult\"at f\"ur Physik, Ludwig-Maximilians-Universit\"at M\"unchen, Scheinerstrasse 1, 81679 M\"unchen, Germany\\
$^{23}$ INAF-Osservatorio Astrofisico di Torino, Via Osservatorio 20, 10025 Pino Torinese (TO), Italy\\
$^{24}$ Dipartimento di Fisica, Universit\`a di Genova, Via Dodecaneso 33, 16146, Genova, Italy\\
$^{25}$ INFN-Sezione di Genova, Via Dodecaneso 33, 16146, Genova, Italy\\
$^{26}$ Department of Physics "E. Pancini", University Federico II, Via Cinthia 6, 80126, Napoli, Italy\\
$^{27}$ INAF-Osservatorio Astronomico di Capodimonte, Via Moiariello 16, 80131 Napoli, Italy\\
$^{28}$ INFN section of Naples, Via Cinthia 6, 80126, Napoli, Italy\\
$^{29}$ Dipartimento di Fisica, Universit\`a degli Studi di Torino, Via P. Giuria 1, 10125 Torino, Italy\\
$^{30}$ INFN-Sezione di Torino, Via P. Giuria 1, 10125 Torino, Italy\\
$^{31}$ Institut de F\'{i}sica d'Altes Energies (IFAE), The Barcelona Institute of Science and Technology, Campus UAB, 08193 Bellaterra (Barcelona), Spain\\
$^{32}$ Port d'Informaci\'{o} Cient\'{i}fica, Campus UAB, C. Albareda s/n, 08193 Bellaterra (Barcelona), Spain\\
$^{33}$ Institute for Theoretical Particle Physics and Cosmology (TTK), RWTH Aachen University, 52056 Aachen, Germany\\
$^{34}$ INAF-Osservatorio Astronomico di Roma, Via Frascati 33, 00078 Monteporzio Catone, Italy\\
$^{35}$ Dipartimento di Fisica e Astronomia "Augusto Righi" - Alma Mater Studiorum Universit\`a di Bologna, Viale Berti Pichat 6/2, 40127 Bologna, Italy\\
$^{36}$ Centre National d'Etudes Spatiales -- Centre spatial de Toulouse, 18 avenue Edouard Belin, 31401 Toulouse Cedex 9, France\\
$^{37}$ Institut national de physique nucl\'eaire et de physique des particules, 3 rue Michel-Ange, 75794 Paris C\'edex 16, France\\
$^{38}$ Institute for Astronomy, University of Edinburgh, Royal Observatory, Blackford Hill, Edinburgh EH9 3HJ, UK\\
$^{39}$ Jodrell Bank Centre for Astrophysics, Department of Physics and Astronomy, University of Manchester, Oxford Road, Manchester M13 9PL, UK\\
$^{40}$ European Space Agency/ESRIN, Largo Galileo Galilei 1, 00044 Frascati, Roma, Italy\\
$^{41}$ ESAC/ESA, Camino Bajo del Castillo, s/n., Urb. Villafranca del Castillo, 28692 Villanueva de la Ca\~nada, Madrid, Spain\\
$^{42}$ University of Lyon, Univ Claude Bernard Lyon 1, CNRS/IN2P3, IP2I Lyon, UMR 5822, 69622 Villeurbanne, France\\
$^{43}$ Institute of Physics, Laboratory of Astrophysics, Ecole Polytechnique F\'ed\'erale de Lausanne (EPFL), Observatoire de Sauverny, 1290 Versoix, Switzerland\\
$^{44}$ UCB Lyon 1, CNRS/IN2P3, IUF, IP2I Lyon, 4 rue Enrico Fermi, 69622 Villeurbanne, France\\
$^{45}$ Departamento de F\'isica, Faculdade de Ci\^encias, Universidade de Lisboa, Edif\'icio C8, Campo Grande, PT1749-016 Lisboa, Portugal\\
$^{46}$ Instituto de Astrof\'isica e Ci\^encias do Espa\c{c}o, Faculdade de Ci\^encias, Universidade de Lisboa, Campo Grande, 1749-016 Lisboa, Portugal\\
$^{47}$ Department of Astronomy, University of Geneva, ch. d'Ecogia 16, 1290 Versoix, Switzerland\\
$^{48}$ INFN-Padova, Via Marzolo 8, 35131 Padova, Italy\\
$^{49}$ Univ Claude Bernard Lyon 1, CNRS, IP2I Lyon, UMR 5822, 69622 Villeurbanne, France\\
$^{50}$ Universit\'e Paris-Saclay, Universit\'e Paris Cit\'e, CEA, CNRS, AIM, 91191, Gif-sur-Yvette, France\\
$^{51}$ School of Physics, HH Wills Physics Laboratory, University of Bristol, Tyndall Avenue, Bristol, BS8 1TL, UK\\
$^{52}$ Aix-Marseille Universit\'e, CNRS/IN2P3, CPPM, Marseille, France\\
$^{53}$ Istituto Nazionale di Fisica Nucleare, Sezione di Bologna, Via Irnerio 46, 40126 Bologna, Italy\\
$^{54}$ University Observatory, Faculty of Physics, Ludwig-Maximilians-Universit{\"a}t, Scheinerstr. 1, 81679 Munich, Germany\\
$^{55}$ Dipartimento di Fisica "Aldo Pontremoli", Universit\`a degli Studi di Milano, Via Celoria 16, 20133 Milano, Italy\\
$^{56}$ INFN-Sezione di Milano, Via Celoria 16, 20133 Milano, Italy\\
$^{57}$ Institute of Theoretical Astrophysics, University of Oslo, P.O. Box 1029 Blindern, 0315 Oslo, Norway\\
$^{58}$ Department of Physics, Lancaster University, Lancaster, LA1 4YB, UK\\
$^{59}$ von Hoerner \& Sulger GmbH, Schlo{\ss}Platz 8, 68723 Schwetzingen, Germany\\
$^{60}$ Technical University of Denmark, Elektrovej 327, 2800 Kgs. Lyngby, Denmark\\
$^{61}$ Cosmic Dawn Center (DAWN), Denmark\\
$^{62}$ Max-Planck-Institut f\"ur Astronomie, K\"onigstuhl 17, 69117 Heidelberg, Germany\\
$^{63}$ Jet Propulsion Laboratory, California Institute of Technology, 4800 Oak Grove Drive, Pasadena, CA, 91109, USA\\
$^{64}$ Mullard Space Science Laboratory, University College London, Holmbury St Mary, Dorking, Surrey RH5 6NT, UK\\
$^{65}$ Department of Physics, P.O. Box 64, 00014 University of Helsinki, Finland\\
$^{66}$ Helsinki Institute of Physics, Gustaf H{\"a}llstr{\"o}min katu 2, University of Helsinki, Helsinki, Finland\\
$^{67}$ NOVA optical infrared instrumentation group at ASTRON, Oude Hoogeveensedijk 4, 7991PD, Dwingeloo, The Netherlands\\
$^{68}$ Universit\"at Bonn, Argelander-Institut f\"ur Astronomie, Auf dem H\"ugel 71, 53121 Bonn, Germany\\
$^{69}$ Aix-Marseille Universit\'e, CNRS, CNES, LAM, Marseille, France\\
$^{70}$ Dipartimento di Fisica e Astronomia "Augusto Righi" - Alma Mater Studiorum Universit\`a di Bologna, via Piero Gobetti 93/2, 40129 Bologna, Italy\\
$^{71}$ Department of Physics, Institute for Computational Cosmology, Durham University, South Road, DH1 3LE, UK\\
$^{72}$ University of Applied Sciences and Arts of Northwestern Switzerland, School of Engineering, 5210 Windisch, Switzerland\\
$^{73}$ Institut d'Astrophysique de Paris, 98bis Boulevard Arago, 75014, Paris, France\\
$^{74}$ Institut d'Astrophysique de Paris, UMR 7095, CNRS, and Sorbonne Universit\'e, 98 bis boulevard Arago, 75014 Paris, France\\
$^{75}$ School of Mathematics and Physics, University of Surrey, Guildford, Surrey, GU2 7XH, UK\\
$^{76}$ European Space Agency/ESTEC, Keplerlaan 1, 2201 AZ Noordwijk, The Netherlands\\
$^{77}$ Department of Physics and Astronomy, University of Aarhus, Ny Munkegade 120, DK-8000 Aarhus C, Denmark\\
$^{78}$ Universit\'e Paris-Saclay, Universit\'e Paris Cit\'e, CEA, CNRS, Astrophysique, Instrumentation et Mod\'elisation Paris-Saclay, 91191 Gif-sur-Yvette, France\\
$^{79}$ Space Science Data Center, Italian Space Agency, via del Politecnico snc, 00133 Roma, Italy\\
$^{80}$ Institute of Space Science, Str. Atomistilor, nr. 409 M\u{a}gurele, Ilfov, 077125, Romania\\
$^{81}$ Departamento de F\'isica, FCFM, Universidad de Chile, Blanco Encalada 2008, Santiago, Chile\\
$^{82}$ Universit\"at Innsbruck, Institut f\"ur Astro- und Teilchenphysik, Technikerstr. 25/8, 6020 Innsbruck, Austria\\
$^{83}$ Institut d'Estudis Espacials de Catalunya (IEEC), Carrer Gran Capit\'a 2-4, 08034 Barcelona, Spain\\
$^{84}$ Institute of Space Sciences (ICE, CSIC), Campus UAB, Carrer de Can Magrans, s/n, 08193 Barcelona, Spain\\
$^{85}$ Satlantis, University Science Park, Sede Bld 48940, Leioa-Bilbao, Spain\\
$^{86}$ Centro de Investigaciones Energ\'eticas, Medioambientales y Tecnol\'ogicas (CIEMAT), Avenida Complutense 40, 28040 Madrid, Spain\\
$^{87}$ Infrared Processing and Analysis Center, California Institute of Technology, Pasadena, CA 91125, USA\\
$^{88}$ Instituto de Astrof\'isica e Ci\^encias do Espa\c{c}o, Faculdade de Ci\^encias, Universidade de Lisboa, Tapada da Ajuda, 1349-018 Lisboa, Portugal\\
$^{89}$ Universidad Polit\'ecnica de Cartagena, Departamento de Electr\'onica y Tecnolog\'ia de Computadoras,  Plaza del Hospital 1, 30202 Cartagena, Spain\\
$^{90}$ Institut de Recherche en Astrophysique et Plan\'etologie (IRAP), Universit\'e de Toulouse, CNRS, UPS, CNES, 14 Av. Edouard Belin, 31400 Toulouse, France\\
$^{91}$ Centre for Information Technology, University of Groningen, P.O. Box 11044, 9700 CA Groningen, The Netherlands\\
$^{92}$ INAF, Istituto di Radioastronomia, Via Piero Gobetti 101, 40129 Bologna, Italy\\
$^{93}$ INFN-Bologna, Via Irnerio 46, 40126 Bologna, Italy\\
$^{94}$ Junia, EPA department, 41 Bd Vauban, 59800 Lille, France}

   \date{Received November 27, 2023; accepted February 4, 2024}

 
  \abstract
   {Dusty, distant, massive ($M_*\gtrsim 10^{11}\,\rm M_\odot$) galaxies are usually found to show a remarkable star-formation activity, contributing on the order of $25\%$ of the cosmic star-formation rate density at $z\approx3$--$5$, and up to $30\%$ at $z\sim7$ from ALMA observations. Nonetheless, they are elusive in classical optical surveys, and current near-infrared surveys are able to detect them only in very small sky areas. Since these objects have low space densities, deep and wide surveys are necessary to obtain statistically relevant results about them. \Euclid will be potentially capable of delivering the required information, but, given the lack of spectroscopic features at these distances within its bands, it is still unclear if it will be possible to identify and characterize these objects.}
   {The goal of this work is to assess the capability of \Euclid, together with ancillary optical and near-infrared data, to identify these distant, dusty and massive galaxies, based on broadband photometry.}
   {We used a gradient-boosting algorithm to predict both the redshift and spectral type of objects at high $z$. To perform such an analysis we make use of simulated photometric observations mimicking the Euclid Deep Survey, derived using the state-of-the-art Spectro-Photometric Realizations of Infrared-selected Targets at all-$z$ (\spr) software.}
   {The gradient-boosting algorithm was found to be accurate in predicting both the redshift and spectral type of objects within the Euclid Deep Survey simulated catalog at $z>2$, while drastically decreasing the runtime with respect to SED-fitting methods. In particular, we study the analog of HIEROs (i.e. sources selected on the basis of a red $H-[4.5]>2.25$), combining \Euclid and \textit{Spitzer} data at the depth of the Deep Fields. 
   These sources include the bulk of obscured and massive galaxies in a broad redshift range, $3<z<7$.
  We found  that the  dusty population at $3\lesssim z\lesssim 7$ is well identified, with a redshift root mean squared error and catastrophic outlier fraction of only $0.55$ and $8.5\%$ ($\HE\leq26$), respectively. 
   Our findings suggest that with \Euclid we will obtain meaningful insights into the role of massive and dusty galaxies in the cosmic star-formation rate over time.}
   {}

   \maketitle
\input{Sections/intro}
\input{Sections/catalog}
\input{Sections/ml}
\input{Sections/discussion}

\begin{acknowledgements}
\AckEC 
\end{acknowledgements}
\bibliographystyle{aa}
\bibliography{references}
\include{Sections/appendix}
\end{document}

%% file: Sections/intro.tex
\section{Introduction}\label{chp:intro}

In the last decades a major effort has been dedicated to the statistical identification of galaxies over a wide range of redshifts. Multi-wavelength observational campaigns (from the X-ray to the radio spectral regime) in the deepest extragalactic fields have allowed the reconstruction of the average properties of various galaxy populations and their evolution. 
A fundamental  result is the measurement of the star-formation rate density (SFRD) of the Universe \citep[e.g.,][]{madaudickinson}. The SFRD reached a peak at  $z\approx1$--$3$ (the so-called ‘‘cosmic noon’’), rapidly declining to the current epoch.
Several works have shown that the fraction of the SFRD obscured by dust, and therefore not accounted for by
optical/UV surveys at $z>2$, is likely not negligible, and increases with redshift at least up to $z\approx5$--$7$ \citep[e.g.,][]{Novak_2017,Gruppioni_2020,Topping2022,Barrufet2023unveiling,Fujimoto2023,algera}.
A comprehensive study of high-redshift galaxies (well beyond cosmic noon) is of fundamental importance for our understanding of the early epochs of galaxy stellar mass assembly.

The classical technique to select sources at $z>3$ relies on their broadband colors, by measuring the drop of brightness caused by the Lyman break (at $912\,\angstrom$ in the rest frame)  and/or the Lyman forest (between $912$ and $1216\,\angstrom$, rest frame). The objects selected are referred to as Lyman-break galaxies (LBGs).
However, while this approach is straightforward to apply, it is also affected by significant incompleteness and contamination; in particular, as a consequence of their redder UV slopes and relative faintness, the LBG selection is known to be notably biased against massive galaxies \citep[$M_*\gtrsim 10^{11}\,\rm M_\odot$;][]{vanDokkum,Bian}. Indeed, various massive, non UV-selected galaxy populations have been detected and spectroscopically confirmed at $z \gtrsim 3$ \citep[e.g.,][]{Daddi2009,Huang2014}.
Among these, optically faint submillimiter galaxies (SMGs, i.e. galaxies discovered at submillimiter wavelengths) have been particularly interesting, as they can be undetectable at high redshift, even with the deepest optical/near-infrared imaging \citep[e.g.,][]{Frayer_2000,wang19,smail}.

These massive and dusty galaxies have low space densities and a remarkable star-formation activity, contributing up to $ \approx20$--$25\%$ of the cosmic star-formation rate density (CSFRD) at $z\approx3$--$5$ \citep{Gruppioni_2020,Talia,Enia22,xiao}.
In particular, the CSFRD estimated for $H$-faint galaxies ($H\gtrsim 26.4,\, < 5\sigma$) in \citet{sun2021extensive} is $\approx 8\%$ of the CSFRD at this epoch \citep{madaudickinson}; the values suggested by \citet{Williams_2019} and \citet{Gruppioni_2020} are approximately $2$ to $3$ times larger.
Despite the importance of these galaxies, given their faintness and non-detections at most wavelengths, most of their physical properties remain highly uncertain except for very few cases with spectroscopic confirmations \citep[e.g.,][]{wang19,Caputi_2021}. 
More recently, some attempts to characterise these galaxy populations have been performed thanks to JWST (e.g., 
\citealt{perezgonzalez,Barrufet2023alma,rodighiero,barro23,bisigello23b}), 
but the areas observed remain small.
It is then clear that mapping the full cosmic star-formation history and understanding the early phases of massive galaxy formation require the study of the star-formation in massive galaxies population to be as comprehensive as possible at $z>3$.

To this end, several color-selection methods have been proposed to identify this optically-faint massive galaxy population. In particular, \citet{Wang2016} present a method based on the $H-[4.5]/J-H$ diagram, enabling a rather clean selection of $z>3$ galaxies (usually called as HIEROs). However, the separation of high-redshift galaxies from low-redshift contaminants remains difficult.
Multi-wavelength observations sampling the spectral energy distribution (SED) of the sources are also widely used to measure their photometric redshifts and physical parameters (such as stellar mass, star-formation rate, stellar ages, extinction, e.g. \citealt{Weaver2022}; \citealt{Laigle16}; \citealt{Ilbert2009}).
The most used technique leveraging this kind of information is template-fitting \citep[e.g.,][]{Benitez2000}, which uses a set of theoretical or empirical SED templates for the estimations.

In the recent years, the large amount of data provided by a wealth of extragalactic surveys has enabled the use of supervised machine learning techniques, in which the mappings between inputs (photometry in different bands) and outputs (redshift or other physical property) are learned through a reference, or training, sample. Their most obvious advantage is the much higher efficiency in memory usage and computational time with respect to SED-fitting techniques, e.g. seconds instead of days when dealing with millions of objects.

Nonetheless, although these new empirical methods were found to outperform even the accuracy of template-based methods \citep{Abdalla2011} and although their use has become very common (e.g., \citealt[][]{Ball_2008}; \citealt{sdss_cnn}; \citealt{Liu2019};  \citealt{Bisigello23a}), there is still no detailed study focusing on distant ($z>3$) galaxies. In fact, the photometric redshift accuracy of these objects is not well determined because of a lack of large enough reference samples with spectroscopic redshifts and the paucity of deep near-IR data.
Furthermore, massive galaxies are rare, and wide fields are needed to obtain statistically relevant results. 

Taking all these issues into account, the upcoming \Euclid Space Telescope \citep{laureijs2011} will open new possibilities for the study of these objects by observing a large area of the sky at near-IR wavelengths. Most of the mission's observations will comprise a wide survey, covering approximately $15\, 000\, \rm deg^2$ down to a $10\sigma$ depth of $24.5\, \mathrm{mag}$ in the visible filter and down to a $5\sigma$ depth of $24\, \mathrm{mag}$ at near-infrared wavelengths. A deep survey two magnitudes deeper than the wide survey will also be conducted, over $50\, \rm deg^2$ in the Euclid Deep Fields \citep{scaramella}.
The Euclid Deep Fields are expected to contain millions of $z > 3$ galaxies and therefore enable studies of early galaxy formation and evolution with unprecedented statistical significance.

Given the limited number of \Euclid photometric bands (a visible band $\IE$ and three near-IR bands $\YE$, $\JE$, $\HE$), efforts are being made to complement \Euclid space-based data with ground-based data in the ultraviolet to visible spectral range \citep{laureijs2011,Ibata2017}. Combined with existing \Spitzer surveys (near-IR light, \citealt{Capak2016}, \citealt{Masters_2019}, e.g. \citealt{Moneti}), these data will provide a comprehensive description of the SEDs of galaxies up to the epoch of reionization.

It is thus essential to assess the capability of the \Euclid filters (combined with the ancillary data) to identify high-redshift galaxies, and in particular the mentioned massive and dusty population, through precise photometric redshifts, and the possibility of characterizing their spectral types.
In this work we propose to perform such a task by using simulated \Euclid photometric observations. We adopt the simulated data from the Spectro-Photometric Realisations of Infrared-selected Targets at all-$z$ (\spr, \citealt{SPRITZ}).
This choice is based on the fact that this phenomenological simulation, which includes both star-forming galaxies and active galactic nuclei (AGN), well reproduces the statistics and the evolution of the far-infrared sources (number counts and luminosity functions), i.e. the dusty massive population.

Since the methodology proposed in this work (i.e. a gradient-boosting algorithm) returns a photometric redshift for each source selected within the simulated light cone, we will present the general performance for the optimized $z>2$ range (where the algorithm is trained).
However, in this paper we will focus our attention on a specific class of objects: the HIEROs (introduced by \citealt{Wang2016} and discussed above). 
They represent an ideal population to test the ability of $Euclid$, combined to ancillary data, in recovering dark galaxies beyond Cosmic Noon.

The structure of the paper is as follows. In Sect. \ref{sec:catalog} the simulated catalog and the data used are introduced. In Sect. \ref{sec:ML} the main methods to achieve our goals are described, followed by the results obtained in Sect. \ref{sec:results}.  Finally, Sect. \ref{sec:discussion} concludes this paper summarizing the main results and the perspectives for the \Euclid mission.

Throughout the paper, we assume a $\Lambda$CDM cosmology with $H_0=70\,\kmsMpc,\, \Omega_{\rm m}=0.27$, $\Omega_\Lambda=0.73$; all magnitudes are in the AB system (\citealt{AB}).

%% file: Sections/catalog.tex
\section{Data}\label{sec:catalog}
\subsection{The \spr simulation}\label{sub:spritz}
\spr \citep[][]{SPRITZ} is a state-of-the-art simulation based on a set of observed galaxy stellar mass functions and luminosity functions, mainly in the IR, derived for different galaxy populations. In this work we consider the version 1.13 of the simulation, with updates the dwarf irregular galaxy stellar mass function used in input and includes CO and [$\ion{\rm C}{II}$] line luminosities, as presented in \citet{Bisigello2022}. Briefly, each simulated galaxy is assigned a unique SED template, taken from a set of 35 empirical templates \citep{Polletta2007,Rieke2009,G10,Bianchi_2018}. These templates are of low-z galaxies, but they represent a good description of galaxies observed by \textit{Herschel} up to $z = 3.5$ \citep{G13} and by ALMA at $z=6$ \citep{Gruppioni_2020}.
A set of empirical and theoretical relations is then used to link each source to its physical properties, such as stellar mass, star-formation rate and AGN contribution. The galaxy populations included in the simulations are spirals, starbursts (SB), ellipticals, dwarf irregulars, AGN and composite AGN. The AGN population includes both type-1 and type-2 AGN, and this classification is based on the optical/UV part of their spectrum. Composite AGN include objects with an AGN component that is not the dominant source of bolometric emission, because they are intrinsically faint (referred to as star-forming AGN, SF-AGN), or because they are extremely obscured by dust (called starburst AGN, SB-AGN). Such galaxy populations are motivated by the variety of galaxies observed by \textit{Herschel} up to $z=3.5$.

The resulting mock catalogs are found to be consistent with a large variety of observations, including the stellar mass versus star-formation rate relation, luminosity functions and number counts from X-ray to radio, demonstrating that this simulation is suitable for making predictions for a set of future surveys operating particularly, but not only, at IR wavelengths. In particular, we highlight that the simulation is in agreement with the IR LF observed at $z\sim6$ \citep{Gruppioni_2020} as well as the [$\ion{\rm C}{II}$] and CO LFs at $z=4$--6 \citep[e.g.,][]{Riechers2019,Loiacono2021,Boogaard2023}. For more details on \spr, we refer to \citet{SPRITZ}.

In this work, we make use of this simulation for forecasting high-redshift galaxy identification in the Euclid Deep Fields.

We note that the results obtained in this work are an optimistic version of the ones that will be obtained from real data.

\subsection{The Euclid Deep Fields simulated catalog}
As reported in Sect. \ref{chp:intro}, the Euclid Deep Survey will observe an area of at least $40\,\rm deg^2$ in 4 filters: 
$\IE$ from the VIS instrument \citep{Cropper16}, and $\YE$, $\JE$, $\HE$ from NISP \citep{Schirmer22},
with $5\sigma$ depths between $26$ and $27.3$ AB magnitudes.

\begin{figure}[h!]
    \centering
    \includegraphics[width=\linewidth]{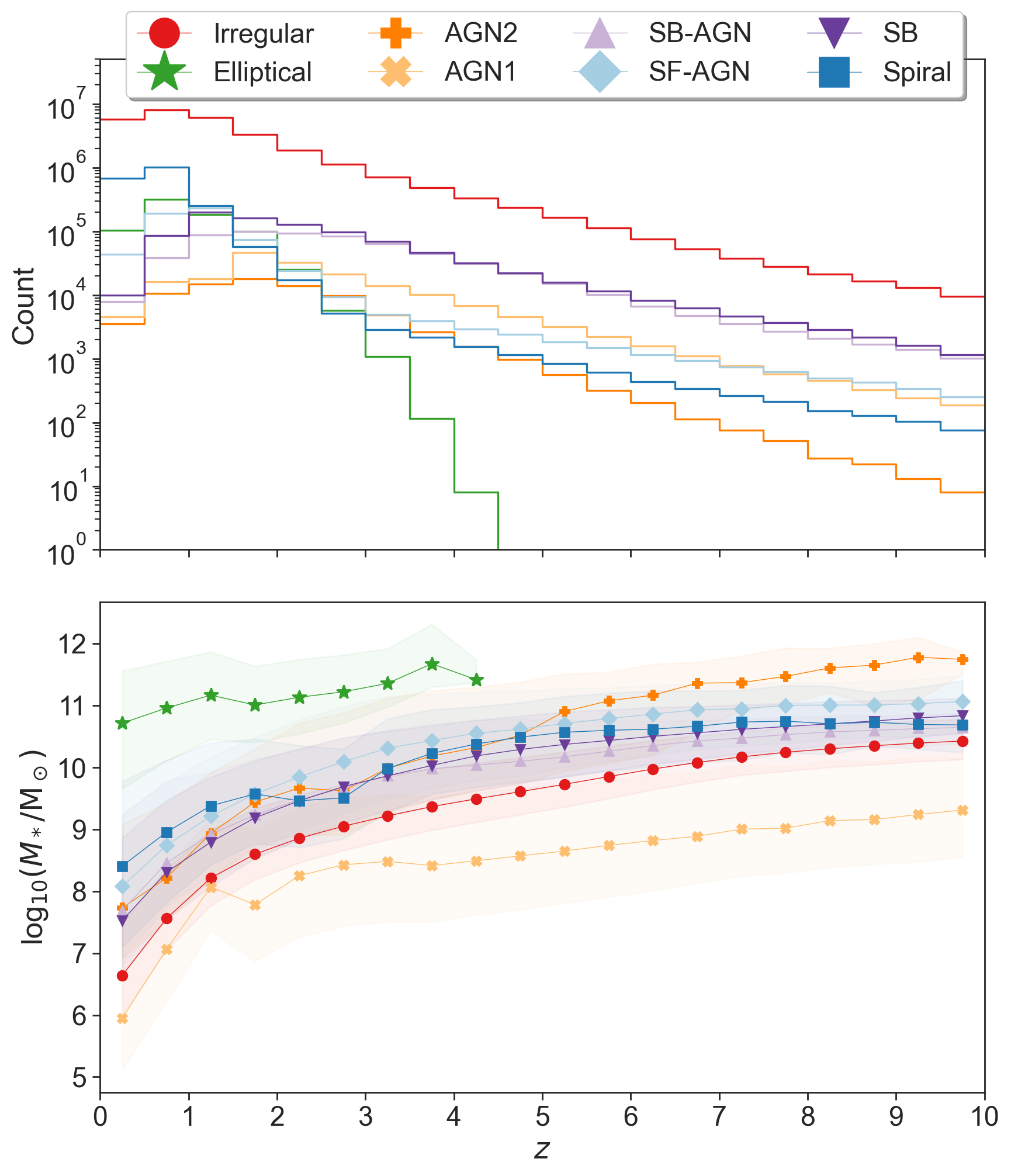}
    \caption{Redshift distribution (top) and stellar mass as function of redshift (bottom) of the simulated galaxies obtained with \spr for the Euclid Deep Survey, color-coded by their SED type (see legend). Markers represent the median value in the redshift bin and shaded areas the corresponding $68\%$ coverage interval. For details about the mass function distribution at different redshift ranges we refer the reader to  \citet{SPRITZ}.}
    \label{fig:Catalogzmass}
\end{figure}

Including galaxies with at least one \Euclid filter with signal-to-noise ratio $ \ge2$, this combination of depth and area results in a simulated catalog with a total of $33\,650\,754$ objects with redshifts from $0$ to $ 10$. 
The redshift distribution and the stellar mass as function of redshift are shown in Fig. \ref{fig:Catalogzmass}. Irregular galaxies with stellar masses below $M_* \approx 10^{11}\,\rm M_\odot$ are expected to dominate the number counts of the survey. At the same time, given the brightness of AGN1, we expect to observe them down to $M_* \approx 10^{8}\,\rm M_\odot$ even at the highest redshifts.

In this simulated catalog, we also include photometry in additional filters, from Rubin and \textit{Spitzer}/IRAC, as these ancillary observations will be available to complement \Euclid space-based data \citep{scaramella}, at least for a fraction of the observed fields. 
It should be noted, however, that the IRAC photometry is affected by source confusion at faint magnitudes. Specifically, at ${\rm IRAC/3.6,\micron}>22$, a de-blending based on the available \Euclid data (\citealt{Moneti}) will be necessary when dealing with real observations.

The expected $5 \sigma $ and $2\sigma$ depths of the \Euclid, Rubin and IRAC bands are reported in Table \ref{table:depths}.
\begin{table}[h]
\caption{Expected $5$ and $2\sigma$ depths of four \Euclid filters, Rubin \textit{ugriz} filters for the Deep Drilling Fields \citep{Foley18} and \textit{Spitzer} IRAC $3.6$, $4.5\, \micron$ \citep{Moneti} filters, as considered in this work.}
\label{table:depths}     
\centering                            
\begin{tabular}{c c c}          
\hline           
Band & $5\sigma$ Depth & $2\sigma$ Depth\\ 
\hline  \hline 
$\IE$ & 28.2&29.2\\
$\YE$ & 26.3&27.3\\
$\JE$ & 26.5&27.5\\
$\HE$ & 26.4&27.4\\
\hline
Rubin/\textit{u} & 26.8&27.8\\
Rubin/\textit{g} & 28.4&29.4\\
Rubin/\textit{r} & 28.5&29.5\\
Rubin/\textit{i} & 28.3&29.3\\
Rubin/\textit{z} & 28.0&29.0\\
\hline
IRAC/3.6$\,\micron$&24.8&25.8\\
IRAC/4.5$\,\micron$&24.7&25.7\\
\hline                                           
\end{tabular}
\end{table}
\subsubsection{Photometric errors}\label{sec:photerr}
In the \spr lightcone used in this work, only the \Euclid photometric bands have an associated observational uncertainty (i.e. the errors on the simulated magnitudes). We perturbed the Rubin and \textit{Spitzer} fluxes to mimic realistic observations. In particular, the Rubin total photometric error has both a systematic and a random contribution and can be written as \citep{Ivezic}:
\begin{align}
    \sigma^2=\sigma_{\rm sys}^2+\sigma_{\rm rand}^2\,,
\end{align}
where $\sigma_{\rm rand}$ is the random photometric error and $\sigma_{\rm sys}$ is the systematic one.
Given that the Rubin telescope is designed to have a systematic error below $0.005\,\rm mag$, we decided to neglect it in this work.
The random photometric uncertainty can be written as function of the magnitude (\citealt{Ivezic}):
\begin{align}
    &\sigma_{\rm rand}^2=(0.04-\gamma)x+\gamma x^2
    \label{eq:error1}\\
    &x\equiv 10^{0.4(m-m_5)}\,,
    \label{eq:error2}
\end{align}
where $m_5$ is the $5\sigma$ depth (see Table \ref{table:depths}) and $\gamma$ is a parameter equal to $0.039$ for the $g,\,r,\,i,$ and $z$ bands and $0.038$ for $u$. The photometry for each galaxy in each Rubin filter was then derived by randomly sampling a Gaussian distribution with mean equal to the true value and standard deviation $\sigma_{\rm rand}$.

Similarly, for the two \textit{Spitzer} bands (IRAC/3.6$\,\micron$ and IRAC/4.5$\,\micron$), we considered errors from \citet{Laigle16} and parametrized them using Eqs. (\ref{eq:error1})--(\ref{eq:error2}) with $m_5$ as in Table \ref{table:depths} and $\gamma=0.038$.

\subsection{Photometric selections}
The first step in our analysis is to exploit the Euclid Deep Field simulated catalog to validate its compatibility with a set of observed photometric diagnostics available from the literature, focusing on high-$z$ galaxies.

As anticipated in the Introduction, the main focus in this work is about the dusty and massive galaxy populations at $3\lesssim z\lesssim7$. For their selection, we rely on the evolutionary $H - [4.5]$ tracks of a set of theoretical galaxy SED templates for $z>3$, assuming the  color cut-off suggested by \citet{Wang2016}:
\begin{align}
    H - [4.5] > 2.25.
    \label{eq:hieros}
\end{align} 
This color  was proposed to select old or dusty galaxies at $z>3$. Objects satisfying this criterion are referred to as HIEROs.
In particular, it was found that almost none of the spectroscopically confirmed Lyman-break galaxies at $z > 3$ satisfies this criterion. These are examples of the types of objects missed by conventional UV/optical surveys that we aim to recover with the future Euclid Deep Survey.
Furthermore, based on the color tracks of theoretical models and their photometric redshifts, HIEROs can be separated in two main classes: 

\begin{align}
    &\text{blue HIEROs,}&H-[4.5]>2(J-H)+1.45\,;
    \label{eq:red}\\ 
    &\text{red HIEROs,}&H-[4.5]\leq 2(J-H)+1.45\,.
    \label{eq:blue}
\end{align}

According to \citet{Wang2016}, blue HIEROs are dominated by normal massive and dusty star forming galaxies at $z \gtrsim 3$. Red HIEROs, instead, include a mix of lower $z\sim2-3$ dusty star forming objects and passive galaxies at $z\sim3-4$. Passive galaxies are expected to be the most massive systems at any cosmic epoch, and are thus relevant for our study.

\begin{figure*}
    \centering
    \includegraphics[width=\linewidth]{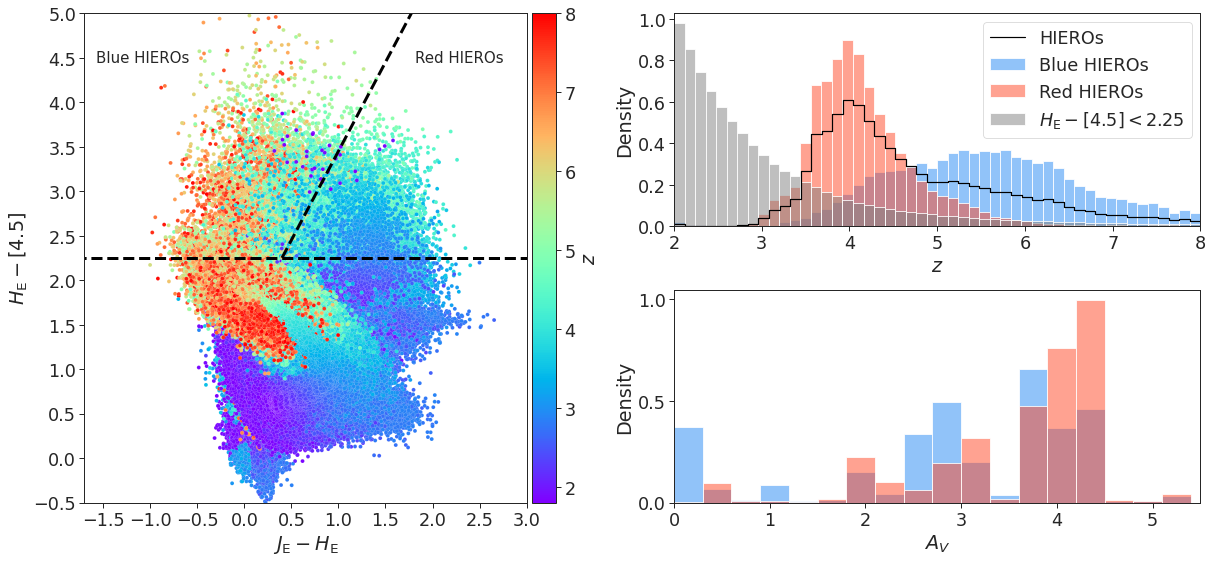}
    \caption{\textbf{Left}. $\HE-[4.5]$ versus $\JE-\HE$ diagram for the Euclid Deep survey simulated catalog ($5\sigma$ depth) color-coded by redshift. \textbf{Top Right}. Redshift distribution of red HIEROs (red), blue HIEROs (blue) and galaxies with $\HE-[4.5]<2.25$ (grey). \textbf{Bottom Right}. Optical extinction distribution of red HIEROs (red) and blue HIEROs (blue). We normalize each histogram to obtain an area equal to unity. }
    \label{fig:wang_cat}
\end{figure*}

Figure \ref{fig:wang_cat} (left panel) shows the $\HE-[4.5]/\JE-\HE$ diagram for the Euclid Deep Survey simulated catalog with $2\leq z\leq 8$ and fluxes brighter than their $5\sigma$ detection limits, colored by redshift. 
To better highlight the different populations, in the top-right panel  we report their redshift distributions: a clear difference arises between HIEROs and other objects, with $99\%$ of HIEROs being at $z>3$ and $43\%$ of non-HIEROs being above the same redshift.

The bottom-right panel shows, instead, the extinction properties for red and blue HIEROs (in terms of $A_V$ distributions). As expected, their red colors are mostly due to the presence of dust, with the bulk of $A_V\sim4$ mag and spanning values up to 5.5 mag.
We note that by selection both blue and red HIEROs include highly extinguished objects \citep{Wang2016}, so it is natural to observe 
consistent $A_V$ distributions (while keeping in mind that blue and red HIEROs peak at different cosmic epochs).

Given that in our simulation red HIEROs dominate the number densities of red sources at $z\sim4$, while blue HIEROs populate the higher redshift queue up to $z\sim 7$--$8$, we will include in the following discussion the study of the overall class of $Euclid$  sources with $H-[4.5]>2.25$, focusing on the $3<z<7$ redshift range (where the bulk of HIEROs lye).

To check the representativeness of our HIERO mock catalog, we have compared the overall number densities with the observations of \citet{Wang2016}, who select HIEROs with [4.5]<24 over an area of 350 arcmin$^2$, detecting a cleaned sample of 285 sources. 
Restricting our selection (i.e. requiring at least four detections at S/N>2  in the considered bands, see Sect. \ref{sub:cleaning})
to the same magnitude limit of \citet{Wang2016}, we identify 155 objects over the same area. However, if we mimic the \citet{Wang2016} selection by simply requiring a 5$\sigma$ detection at 4.5$\mu$m, and  [4.5]<24, the number of predicted HIEROs increases to 425 (over 350 arcmin$^2$). 
The variance on the predicted space densities, related to the assumed selection function, shows that the our mock catalog provides a statistical sampling of the HIERO population consistent with the real world.

\begin{figure}
    \centering
    \includegraphics[width=\linewidth]{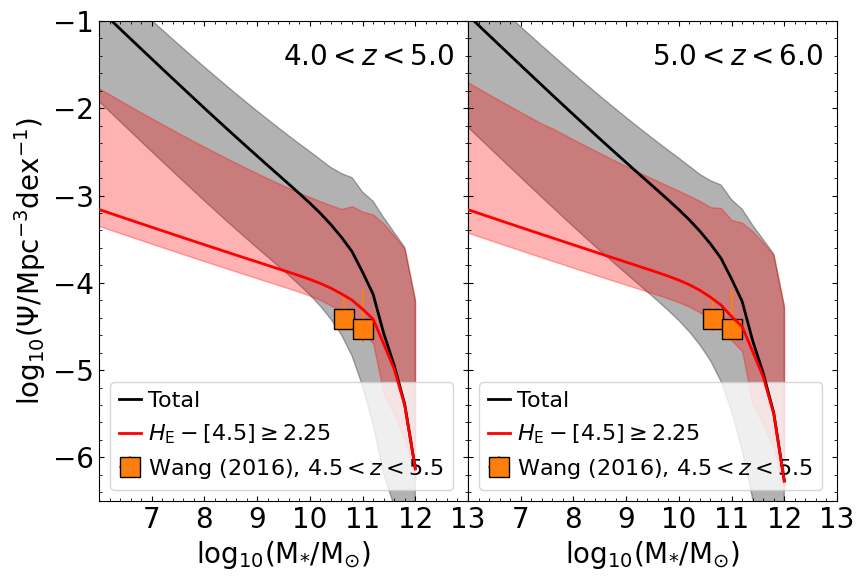}
    \caption{Contribution of HIEROs to the galaxy stellar mass function in \spr at $z>4$ (red line). The grey and red shaded areas include the errors of the initial luminosity functions used to derive simulated galaxies and the uncertainties due to the high-$z$ ($z>3$) extrapolation. The observed stellar mass function for HIEROs measured by \citet{Wang2016} at $4.5<z<5.5$ is reported in both panels as filled orange squares.}
    \label{fig:MF_hieros}
\end{figure}

To understand the role of HIEROs in the stellar mass assembly, we report in Fig. \ref{fig:MF_hieros} (black lines) the overall stellar mass function for galaxies in the redshift range $4<z<6$ from \spr,  splitted in two $z$ bins.
The  mass function limited to HIEROs is also reported (red lines),
showing the dominant contribution of this population to the massive end of the stellar mass function at $z>4$. 
Both at $z=4.5$ and $5.5$, HIEROs constitute approximately $93\%$ of the galaxies with $M_*>10^{11.5}\,\rm M_\odot$ and $32\%$ of the galaxies with $M_*>10^{10.5}\,\rm M_\odot$.
The Figure includes the observed mass function of HIEROs at $z\sim4-5$ from \citet{Wang2016}, confirming the consistency of our model with the observations, within the uncertainties.

\begin{figure}
    \centering
    \includegraphics[width=\linewidth]{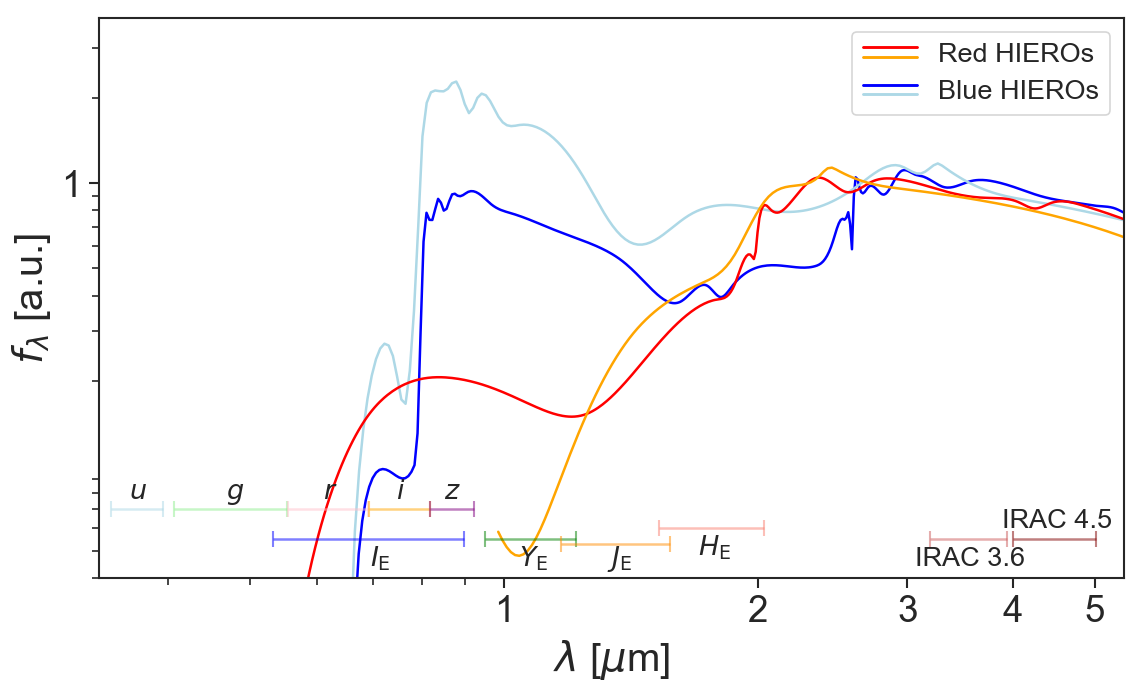}
    \caption{The four most numerous SED templates used in \spr for the generation of red HIEROs (red and orange) as considered at $z=4$, and blue HIEROs (blue and light blue) at $z=5.5$. The flux is in arbitrary units. We also show the wavelength coverage of the photometric bands considered in this work.}
    \label{fig:SED_hieros}
\end{figure}

Finally, in Fig. \ref{fig:SED_hieros} we show the most representative SED templates in \spr for this population of galaxies, specifically the four most numerous templates used to generate red and blue HIEROs.

%% file: Sections/ml.tex
\section{Methods}\label{sec:ML}
In this section, we introduce the methods and the metrics that we used to estimate the capability of the future \Euclid Space Telescope to identify high-redshift galaxies.
\subsection{Gradient-boosted trees}\label{sub:XGB}
We have considered a gradient-boosting approach to independently predict both the redshifts and the SED types, based on the observed fluxes in different bands. This corresponds to a regression and a classification problem, respectively.

Other methods have been proposed in the literature to classify different galaxy populations in \Euclid (e.g. \citealt{Bisigello2020}, \citealt{Hump22}) or to derive photometric redshifts (e.g., \citealt{Deprez2020}, Bisigello et al. in prep.), but none of these is focused on high-$z$ galaxies.

\begin{figure}
    \centering
    \includegraphics[width=0.8\linewidth]{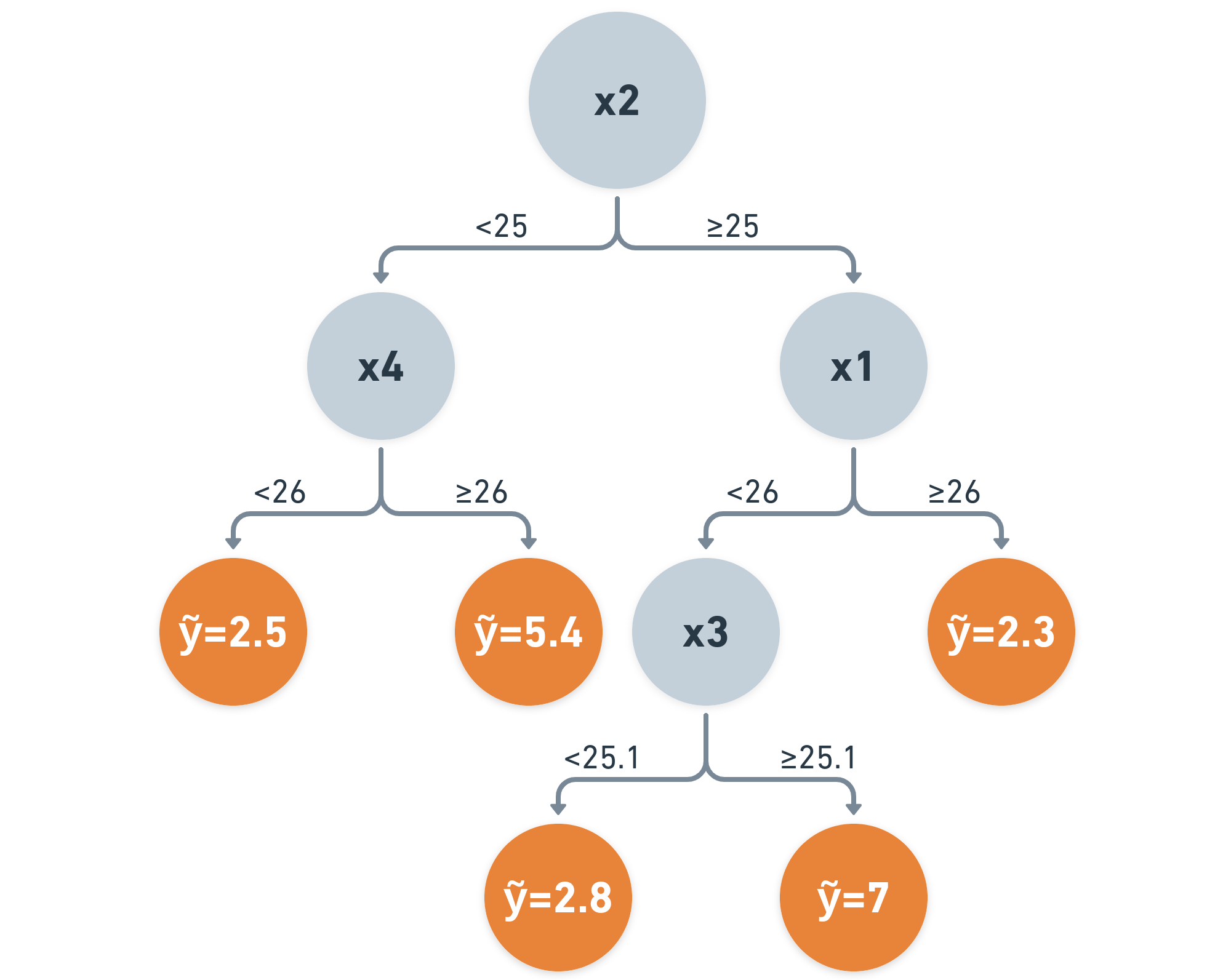}
    \caption{Decision tree model. Grey nodes are called \textit{internal nodes}, while orange ones, which represent the tree predictions, \textit{leaf nodes}. Features are indicated with $x1,\,x2,\,x3,\,x4$ and predictions with $\tilde{y}$. Every internal node is labeled with an input feature. The arcs coming from a node labeled with an input feature (for example $x4$) to a leaf node (orange) are labeled with each of the possible values of the target feature (for example $\tilde{y}=2.5$ or $\tilde{y}=5.4$); otherwise the arc leads to a subordinate decision node on a different input feature. For example the arcs starting from the node labeled $x2$ lead to decisions based on the value of $x4$ and $x1$.}
    \label{fig:DT}
\end{figure}

The method proposed in this paper is a type of ensemble algorithm, in which the relationships between the features $\mathbf{x}$ and the target variables $y=f(\mathbf{x})$ are learned by sequentially fitting new models: new decision trees (a representation of the decision tree model is shown in Fig. \ref{fig:DT}) are constructed to be maximally correlated with the negative gradient of the loss function associated with the ensemble.
The loss function is chosen according to the task, while structural and learning parameters (referred to as hyperparameters) have to be tuned in a data-driven fashion.

The algorithm used in this work is implemented in the software package {\tt XGBoost} (\citealt{XGBoost}). Here we briefly summarise its main features while for more details we refer to the paper in which the library is presented or the online documentation.\footnote{\href{https://xgboost.readthedocs.io}{https://xgboost.readthedocs.io}}
There are four main reasons to use gradient-boosted trees, and in particular {\tt XGBoost}, in this work:
\begin{enumerate}
    \item Execution speed: Generally, {\tt XGBoost} is faster than backpropagation-based models (e.g. neural networks), and even than some other gradient-boosting implementations (\citealt{GBBenchmark}). In addition, and of particular interest for galaxy identification, its support for hardware acceleration makes the speed difference with any SED-fitting algorithm very advantageous (see Sect. \ref{sec:hyperparam});
    \item Model performance: Gradient-boosting dominates structured and tabular data sets on both regression and classification predictive modelling problems. As evidence, it is the go-to algorithm for competition winners on Kaggle\footnote{\href{https://www.kaggle.com/}{Kaggle} is a community of data scientists that allows users to find and publish data sets, explore and build models in a web environment and participate in data science and machine learning competitions (with prize).};
    \item Decision-based splits: In the decision tree model, the splitting occurs at certain thresholds for every feature (see Fig. \ref{fig:DT}). This makes the model more robust to outliers and to the presence of only upper or lower limits in certain attributes, because it does not make any difference how far a point is from such thresholds;
    \item Missing values handling: The data set used in this work includes missing data (NaNs) in one or more of the features. Traditionally, to handle NaNs one can specify a fixed value to replace missing numbers, or impute them with either the mean or the median of that feature. It is obvious that this approach might not always be the best choice. {\tt XGBoost} enhances the function class to learn the best way to handle missing values: the idea is to learn a ‘‘default direction’’ for each node and guide the sample with missing values along the default directions. This approach can be seen as an implicit way of imputing missing numbers.
\end{enumerate}
We might also have considered other recent gradient-boosting implementations, for example {\tt CatBoost} (\citealt{CatBoost}) or {\tt LightGBM} (\citealt{lightgbm}), which should perform very similarly to {\tt XGBoost}. However, a systematic comparison of different gradient-boosting libraries is beyond the scope of this work. 
We opted for {\tt XGBoost} due to its longer presence in the field, which has allowed for a more mature development of features and a wider user community.
\subsubsection{{\tt XGBoost} hyperparameters}
A hyperparameter is a parameter whose value is used to control the learning process, which, as a consequence, cannot be learned, but has to be set by the user by evaluating the machine performance while varying its value.

The most impactful {\tt XGBoost} hyperparameters are as follows:
\begin{itemize}
\item The number of estimators refers to the number of gradient-boosted trees fitted during the learning process. Larger values lead to more complex models, which, however, are more prone to overfitting\footnote{Overfitting is the production of an analysis corresponding too closely to a particular data set, and might thus fail to predict new observations reliably.}.
\item The learning rate is the rate at which new trees are added to the ensemble. Lower values lead to a slower addition of new trees, thus preventing (or at least slowing down) overfitting.
\item The maximum depth refers to that of a tree. Increasing this value will make the model more complex and more likely to overfit.
\item $\gamma$ is defined as the minimum loss reduction required to make a further partition on a leaf node of the tree. The larger $\gamma$ is, the more conservative the algorithm will be.
\item $\lambda$ is the L2 (squared norm) regularization term on the weights. Increasing this value will make the model more conservative.
\item The column sample by tree is the subsample ratio of columns when constructing each tree. Subsampling occurs once for every tree constructed.
\end{itemize}
The values used in this work and the optimization method used to derive them are reported in Sect. \ref{sec:hyperparam}.

\subsection{Data preprocessing and feature engineering}\label{sub:preprocessing}
Data preprocessing is a fundamental step in machine learning, since the quality of data greatly impacts the capability of a model to learn.
Therefore, before feeding the data to the machine learning model and tuning its hyperparameters, we preprocessed it by performing feature engineering and data cleaning.

To begin, the magnitude cut applied to the \Euclid simulated lightcone is set to a depth of $2\sigma$ and we define magnitudes larger than their $2\sigma$ limit (see Table \ref{table:depths}) as missing numbers.

In the following analysis, we consider only objects at $2\leq z\leq 8$. This selection yields a data set with $6\,304\,179$ galaxies.
We have not considered galaxies at $z<2$ as the official photo-$z$ pipelines are very well optimized for identifying low-redshift galaxies (see, for e.g. \citealt{Bisigello23a}; \citealt{Deprez2020}). By utilizing them, we then assume that we will be able to have a clean selection of high-redshift objects.

To test the validity of this assumption, an analysis addressing potential contamination from low-redshift sources in the selection of HIEROs candidates is presented in Appendix \ref{sec:contaminants}. In the same appendix we also assess the contamination by brown dwarfs.

\subsubsection{Features engineering}\label{sub:FI}
In this work, each element in the data refers to the simulated photometry of a particular galaxy.
To provide more information for the objects, some derived features are also included: the more that is known about the SED of a galaxy, the better the inference will be. The number of features provided in the data set is limited to $11$ (four \Euclid, five Rubin and two IRAC bands, see Table \ref{table:depths}), therefore we decided to include some additional features to add more information for the training. In particular, we included the following features.
\begin{itemize}
    \item Differences: pairwise (without permutation) differences of the magnitudes;
    \item Ratios: pairwise ratios between magnitudes without permutation. Even though they have no physical meaning, they are used because we empirically found that they help the training, increasing (albeit slightly) the performance;
    \item Errors: parametric photometric errors associated with each band, as given by Eq. (\ref{eq:error1}) (applied to the perturbed magnitudes). This parametrization is also applied for the \Euclid magnitudes even tough their uncertainties are provided in the catalog, as these are computed analytically starting from the true flux, unaffected by photometric errors.
\end{itemize}
This process generates a total of $132$ features, whose importance is reported and discussed in Appendix \ref{app:fi}.
\subsubsection{Data cleaning}\label{sub:cleaning}
To have a more reliable set of measurements, only objects detected (i.e. ${\rm S/N}>2$) in at least four bands are used for estimating photometric redshifts; their counts for different $z$ ranges are reported in Table \ref{table:7bandscounts} and the fraction of detections per band (computed after the cleaning procedure) in Table \ref{table:detections_per_band}. The choice of going to such a low  ${\rm S/N}$ is explained in Sect. \ref{sub:training_size}. We remind the reader that the starting catalog contains galaxies with at least one detection in a \Euclid filter.

This cleaning procedure removes roughly $18\%$ of the initial data, yielding a total of $5\,174\,988$ galaxies. To show how the redshift distribution is not strongly affected by the cleaning performed, the percentage reduction of objects in different redshift ranges is also reported in Table \ref{table:7bandscounts}.
\begin{table*}
\setlength{\tabcolsep}{5pt}
\caption{Counts per squared degree for different redshift intervals and spectral types for objects detected at least in four bands. In parentheses we indicate the percentage of objects lost during data cleaning, which does not show a strong dependence on redshift. We lose 100\% of elliptical galaxies at $4\leq z<5$, but the starting catalog contained only eight of them in this redshift range. We also present the counts for HIEROs, which are also only marginally affected by our data cleaning process.}
\label{table:7bandscounts}      
\centering                                      
\begin{tabular}{c c c c c c c c c c c}      
\hline              
$z$ & Total &  Spiral & SB & SF-AGN & SB-AGN & AGN1 & AGN2 & Elliptical & Irregular & HIEROs\\  
\hline\hline                                
\multirow{2}{*}{$2\leq z<3$}&73906 & 508 & 4612 & 784 & 3498 & 1267 & 480 & 598 & 62159 & 152\\
\vspace{1mm}&(17.08\%)&(9.47\%) & (18.32\%) & (6.08\%) & (20.19\%) &  (5.22\%) & (18.22\%) & (22.92\%) & (17.13\%) & (27.17\%) \\

\multirow{2}{*}{$3\leq z<4$}&30205 & 111 & 2447 & 205 & 2195 & 568 & 165 & 10 & 24504 & 1107 \\
\vspace{1mm}&(17.38\%)&(11.1\%) & (14.97\%) & (7.05\%) & (18.11\%) &   (6.0\%) & (10.94\%) &  (67.0\%) & (17.87\%) & (7.81\%)\\

\multirow{2}{*}{$4\leq z<5$}&14027 & 62 & 1179 & 125 & 1092 & 258 & 57 & 0 & 11253 & 1979\\
\vspace{1mm}&(19.31\%)&(7.54\%) & (12.41\%) &  (5.9\%) & (17.03\%) &  (8.89\%) &  (9.88\%) & (100\%) & (20.61\%) & (4.55\%)\\

\multirow{2}{*}{$5\leq z<6$}&6554 & 34 & 595 & 78 & 526 & 120 & 20 & 0 & 5182 & 1443\\ 
\vspace{1mm}&(22.45\%)&(6.36\%) & (12.66\%) & (6.66\%) & (16.26\%) & (10.72\%) &  (9.46\%) &   (0\%) & (24.54\%) & (3.57\%)\\

\multirow{2}{*}{$6\leq z<7$}&3144 & 18 & 311 & 49 & 255 & 60 & 7 & 0 & 2444 & 801\\
\vspace{1mm}&(21.49\%)&(7.55\%) & (13.79\%) & (5.92\%) & (10.58\%) &  (10.7\%) & (10.48\%) &   (0\%) & (23.91\%) & (3.94\%)\\

\multirow{2}{*}{$7\leq z<8$}&1540 & 11 & 175 & 31 & 125 & 30 & 3 & 0 & 1166 & 356\\
&(25.85\%)&(10.13\%) & (16.18\%) &  (8.7\%) & (19.66\%) &  (9.79\%) & (16.67\%) &   (0\%) &  (28.5\%) & (17.88\%)\\
\hline                                             
\end{tabular}
\end{table*}
\begin{table*}
\caption{Fraction of detections, i.e. ${\rm S/N}>2$, per considered band for objects detected at least in four bands.}              
\label{table:detections_per_band}   
\centering                                 
\begin{tabular}{c c c c c c c c c c c}      
\hline       
Rubin/\textit{u}&Rubin/\textit{g}&Rubin/\textit{r}&Rubin/\textit{i}&Rubin/\textit{z}&$\IE$& $\YE$& $\JE$& $\HE$&
IRAC/$3.6\,\micron$&IRAC/$4.5\,\micron$\\

16.4\%&75.8\%&94.6\%&96.2\%&93.2\%&
94.5\%&46.3\%&71.6\%&87.2\%&
60.1\%&61.6\%\\
\hline                                             
\end{tabular}
\end{table*}
In the following photometric redshift estimation procedure, for bands with missing numbers (which replaced magnitudes fainter than their $2\sigma$ detection limits), $2\sigma$ magnitude lower limits are used (see Table \ref{table:depths}).
Missing colors and ratios of magnitudes are also replaced with the $2\sigma$ magnitudes lower limits of the missing band (if only one) instead of their upper/lower limit, or left missing (if both magnitudes are missing).

These choices were taken in order to avoid the contamination of detected colors and ratios with other lower/upper bounded ones; they furthermore provided slightly better performance. Some clarifying examples are shown in Table \ref{table:NaN_Handling}.

\subsection{Performance metrics}
Three metrics are used to evaluate the redshift prediction performance ($m$ indicates the number of objects in the sample, $z$ the true redshift and $\Tilde{z}$ the model-predicted redshift):
\begin{center}
\begin{tabular}{p{0.32\linewidth}p{0.58\linewidth}}
    Root Mean Squared Error  & $\rm RMS=\sqrt{\frac{1}{m}\sum_{i=1}^m (z_i-\Tilde{z}_i)^2}\,;$\\
    \\
    Bias& $\langle\Delta z\rangle=\frac{1}{m}\sum_{i=1}^m (\Tilde{z}_i-z_i)\,;$\\
    \\
    Normalized Median Absolute Deviation& ${\rm NMAD}=1.48\, {\rm median} \bigg(|z-\Tilde{z}|/(1+z)\bigg)\,;$\\
    \\
    Catastrophic OutLier Fraction& ${\rm OLF} = \frac{1}{m}\big| \big\{\Tilde{z}\,:\,|z-\Tilde{z}|/(1+z)>0.15\}\big|\,.$
\end{tabular}
\end{center}
For spectral type classification, to evaluate the model we use a simple accuracy metric:
\begin{gather*}
   \quad\quad\quad\quad\,\quad\text{Accuracy} = \frac{\text{Number of correct predictions} }{\text{Total number of predictions}}\,.
\end{gather*}
\subsection{The training set} \label{sub:training_size}
To train and evaluate the model, the data set is divided into two subsets.
The first one, of size $N_{\rm train}$, is used to fit the model and is referred to as the \textit{training set}. The second, of size $N_{\rm test}$, is not used to train the model; instead, its input elements are provided to the model to make predictions. This second subset is referred to as \textit{test set}. The objective of such division is to estimate the performance of the machine learning model on new data. This procedure is called train-test split and depends on the percentage size of the training and test sets with respect to the initial data set.

To ensure the model is trained effectively, in real-world observations a training set must be built with the most robust measurements and accurate redshift estimations available. The redshifts in a training set are thus required to be estimated spectroscopically or with very reliable photometry derived from a larger number of bands than those that will be available with \Euclid + ancillary data.
Consequently, as detailed in Sect. \ref{sub:cleaning}, we decided to extend our analyses to objects with ${\rm S/N}>2$, as in real observations these objects will be supplemented with additional information that compensates for such a low ${\rm S/N}$.

While in practical applications one typically has limited control over the size of the training set $N_{\rm train}$, when forecasting future surveys observations it is useful to assess what dimension of the training set is required to obtain a given prediction performance. In this first part, the minimum training set size required to obtain good test performance is found by performing a baseline {\tt XGBoost} regression on different numbers of training points $N_{\rm train}$ and then comparing the results obtained.

\begin{figure*}
    \centering
    \includegraphics[width=\linewidth]{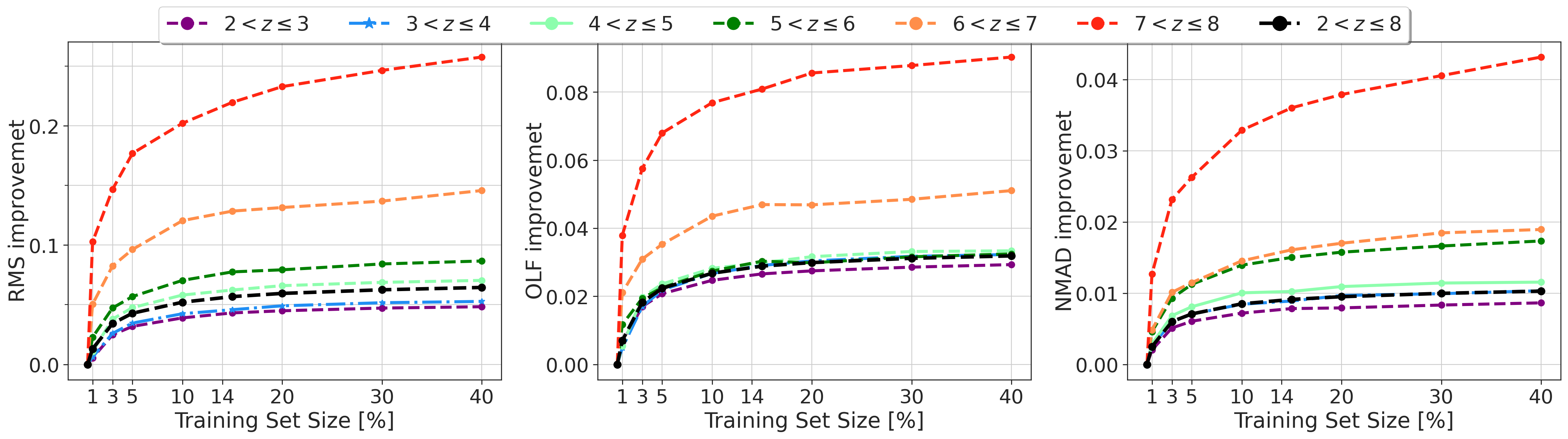}
    \caption{Photometric redshift prediction performance improvement (as defined in Footnote \ref{fnlabel}) with respect to the metrics with the initial training set size ($0.5\%$ of the total number of galaxies), as a function of the training set size, evaluated on a sample of $1\,000\,000$ test galaxies.
    From left to right, we plot: $\rm RMS$, $ \rm OLF$, $\rm NMAD$ for different redshift intervals, as indicated in the legend.}
    \label{fig:tt_res}
\end{figure*}

Figure \ref{fig:tt_res} shows the $\rm RMS$, $\rm OLF$ and $\rm NMAD$ improvement (evaluated on a test set of $1\,000\, 000$ galaxies subdivided into different redshift ranges) for different training set sizes\footnote{The improvement of a metric for a training set of size $N_{\rm train}$, with respect to the initial training set size one, $M_0$, is defined as $M_0-M(N_{\rm train})$, where $M(N_{\rm train})$ is the metric evaluated after training with $N_{\rm train}$ data points.\label{fnlabel}}, as a function of $N_{\rm train}/(N_{\rm train}+N_{\rm test})$.
Analyzing such curves, the following conclusions are clear:
\begin{itemize}
    \item All three performance metrics reported increase as $N_{\rm train}$ increases, as expected, but they begin to flatten for large $N_{\rm train}$ values.
    In general, the size of the training set required to obtain a certain precision depends on the diversity built into the data set: for photometric redshifts it depends on the redshift range, meaning that the narrower the redshift range, the smaller the $N_{\rm train}$ required;
    \item The performance gain is in every case larger for higher redshift objects with respect to lower redshift ones; for example, the $\rm RMS$ improvement for objects at $6<z\leq7$ becomes more than five times larger than the one for $z\leq3$ galaxies.
    \end{itemize}
The number of detected galaxies at $z<4$ is about four times the number of higher-$z$ galaxies (see Table \ref{table:7bandscounts}). This means that larger training sets are important to constraint the redshifts of rare objects: the majority of galaxies in the data set (and, consequently, in the training set) are at low redshifts, so the training process gives more weight to the predictions for galaxies at low redshifts with respect to the other galaxies. Furthermore, for machine learning models, high-$z$ objects are more difficult to identify than lower-$z$ ones, given their larger photometric uncertainties.

To conclude, all the curves reported in Fig. \ref{fig:tt_res} show an onset of a plateau at an $N_{\rm train}$ corresponding to roughly $10\%$ of the data size. This $10\%$-$90\%$ train-test split will then be used in the following analyses.
\subsection{Hyperparameters tuning and time performance}\label{sec:hyperparam}
A Bayesian hyperparameters optimization with cross-validation is then run on a training set large $10\%$ of the total data size. The hyperparameters delivering the lowest $\rm RMS$ thus found are reported in Table \ref{table:hyperparam}.
\begin{table}[h]
\caption{Best {\tt XGBoost} hyperparameters.}            
\label{table:hyperparam}   
\centering                                 
\begin{tabular}{c c}      
\hline       
Number of estimators&$530$\\
Learning rate&$0.01$\\
Max depth&$12$\\
$\gamma$&$0$\\
$\lambda$&$7.7\times 10^{-5}$\\
Column sample by tree&$0.6$\\
\hline                                             
\end{tabular}
\end{table}

Timed on a workstation with a $2.20\,\rm GHz$ Intel Xeon CPU and a $16\,\rm GB$ Tesla P100-PCIE GPU, an {\tt XGBoost} regressor with these hyperparameters takes $70$ seconds to train on a data set with $517\,498$ samples and 132 features, and $40$ seconds to estimate the redshift for the $~4.6$ million galaxies in the test set.

To demonstrate the difference in speed with SED-fitting methods, performing this operation with {\tt LePHARE} (\citealt{LePHARE}) on similar hardware requires approximately $0.23$ seconds per object (considering $463\,680$ different combinations of SED template, redshift, age and dust extinction). Estimating the photometric redshift for all the galaxies in the test set would then need approximately $12$ days.
\section{Results}\label{sec:results}
During the testing phase, we set a higher significance threshold by considering only observations with at least one band having ${\rm S/N} > 5$. This additional criterion, in conjunction with the requirement of at least four detections at ${\rm S/N} > 2$ (see Sect. \ref{sub:cleaning}), aimed to minimize inclusions of spurious objects such as noise spikes, halos, and artefacts that would result from real observations.
This removes roughly 21.728\% of the initial test data $N_{\rm test}$, yielding a total of $3\,645\,481$ galaxies.
The results reported thereafter are thus relative to this data set.
\subsection{Photometric redshifts}\label{sec:photoz}
\begin{figure*}[h!]
    \centering
    \includegraphics[width=\linewidth]{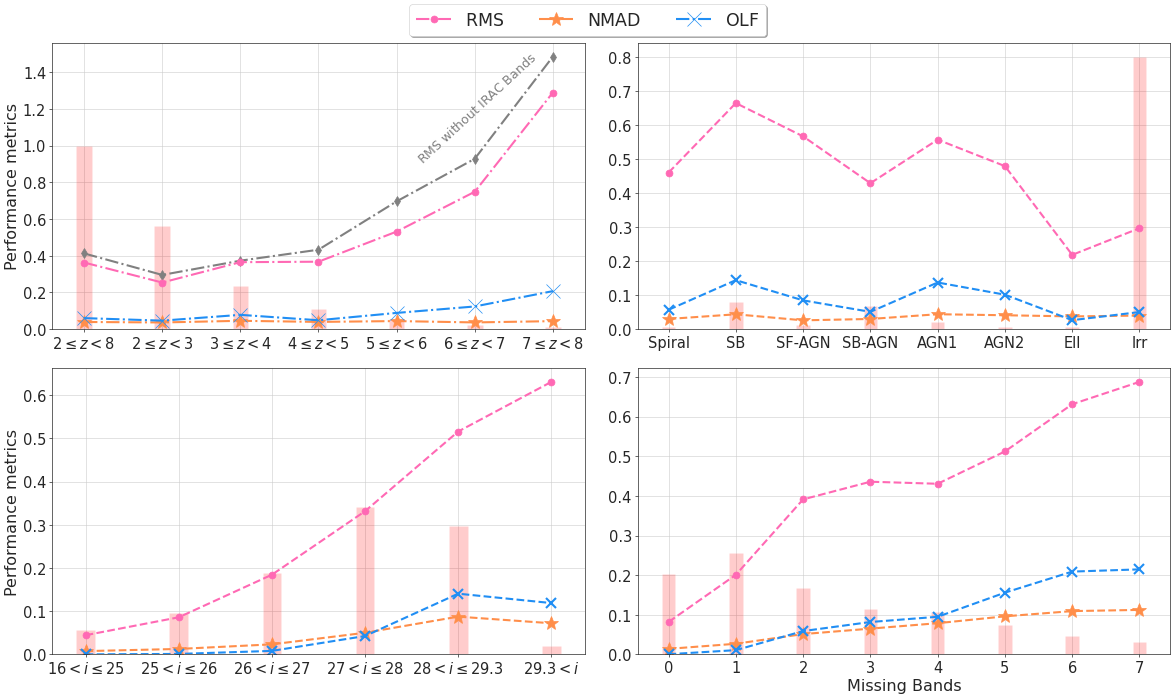}
    \caption{{\tt XGBoost} test set $\rm RMS$, $\rm OLF$ and $\rm NMAD$ (starting from the upper left panel, clockwise) for different redshift intervals, SED types, number of missing bands and $i$-band magnitudes. The grey curve shows the $\rm RMS$ obtained by removing the two IRAC bands from the {\tt XGBoost} input. The vertical pink bands indicate the fraction of objects belonging to the group with respect to the test set size. For details refer to Table \ref{table:1090res}.}
    \label{fig:total_evals}
\end{figure*}

The photo-$z$ performance for the test set is summarized in Table \ref{table:1090res} and shown in Fig. \ref{fig:total_evals}.
This figure clearly illustrates the trend of the performance metrics:
\begin{enumerate}
    \item Top left panel. The errors increase with redshift, as expected and noted in Sect. \ref{sub:training_size};
    \item Bottom panels. The precision of the predicted redshift decreases for fainter objects, as these are usually found at higher $z$ and they come with detections in fewer bands; this practically means the input features vectors contain less information to learn and infer from.
    \item Top right panel. The lowest errors ($\rm RMS=0.22$) are obtained for elliptical galaxies, and the largest for AGN1, SF-AGN and starbursts ($\rm RMS>0.5$). This behaviour may be related to their variability\footnote{Data variability refers to how spread out a set of data is.} and (for SB) their number density evolution with redshift: while ellipticals quickly drop from $z=2$ to disappear at $z\gtrapprox3.5$, SBs are observed at any redshift;
\end{enumerate}
Furthermore, by removing the IRAC bands from the input features, we note (Fig. \ref{fig:total_evals}, top left panel) how mid-IR data improves redshift estimation especially for $z\gtrsim4$ galaxies.
\subsubsection{Comparison with previous results}
To give context to the results obtained, we compare them to previous photometric redshift performance, in a similar redshift range, as reported in \citet[][hereafter W22]{Weaver2022}. There, the precision of the photometric redshifts obtained with {\tt LePHARE} using 39 bands, and included in the COSMOS2020 catalog, is assessed against spectroscopic ones over the COSMOS field ($0<z<6$).

The photometric catalogs created in COSMOS, with their rich multi-wavelength coverage, have for years constituted the state of the art reference to predict the quality of photo-$z$s. Within the Euclid Collaboration \citep{Deprez2020}, this assessment has been done using the COSMOS2015 catalog (\citealt{Laigle16}).
Improving on previous releases, the COSMOS2020 catalog features significantly deeper optical, infrared, and near-infrared data and thus gains almost one order of magnitude in photometric redshift precision compared to its predecessor COSMOS2015.

Comparing the results reported in \citetalias{Weaver2022} and those obtained in this work using a gradient-boosting approach on the Euclid Deep Survey simulated catalog, we find that even in the faintest $25 < i < 27$ bin, gradient-boosting provides a lower $\rm NMAD$ ($0.019$ versus $0.044$), and also a lower $\rm OLF$ ($0.005$ versus $0.204$).
Overall, the performance of the {\tt XGBoost} method presented in this work is comparable to the performance of previous SED-fitting at each magnitude range, but our process is based on a notably smaller number of filters as inputs (i.e. 11 instead of 39). 

It is important to remind the reader that our results are derived from simulated data, which is a simplified representation of reality, and therefore may not reflect actual performance accurately and could potentially overestimate it. The aim of this comparison is therefore to show that our performance is in line with previous studies without stressing much the one-to-one direct comparison.

To conclude this photo-$z$ section, we estimate that in the range $ z> 6 $ the fraction of contaminants  is relatively low, at $ 12 \% $, while the completeness is around $72\%$.
For comparison, \citet{VanMierlo}, using the same set of photometric bands (\Euclid+ Rubin + \textit{Spitzer}) and estimating with {\tt LePHARE} the photometric redshifts of mock galaxies created from the UltraVISTA ultra-deep survey, obtained at $z>6$ a contamination fraction of $12\%$ for bright UltraVISTA-like galaxies, with a $z>6$ completeness of $95\%$. For fainter sources ($25.3 \leq \HE < 27.0$), contamination is more prevalent at $35\%$, with a $z>6$ completeness of $88\%$.
Our smaller completeness and contaminant fractions are mostly due to the unbalanced redshift distribution in the data set (and consequently in the training set), as pointed out in Sect. \nolinebreak\ref{sub:training_size}. In fact, objects at $z>6$ are systematically placed at lower redshifts, as their average offset $\Tilde{z}-z$ is $\approx-0.483$, while for objects at $z<6$ it is $\approx0.001$.
However, this paper focuses mainly at $3<z<7$ , where the {\tt XGBoost} performance is better, with a completeness of $89.3\%$ and a contaminant fraction of $7.73\%$.
\subsection{SED type classification}\label{sect:SED_class}
We also performed an independent experiment, applying the same methodology with the aim of recovering the classification of the spectral class of each source. The prior knowledge on the redshift derived as in the previous sections is ignored here.

When used as a classifier, when a feature vector is fed to it, the {\tt XGBoost} output is a vector of probabilities, in this case with eight entries, each one corresponding to an SED type. The predicted class is thus the one corresponding to the maximum value.
Each simulated galaxy is assigned an SED template taken from a set of 35 templates, divided into eight spectral types, as reported in Sect. \ref{sub:spritz}.

\begin{figure}[h!]
    \centering
    \includegraphics[width=.9\linewidth]{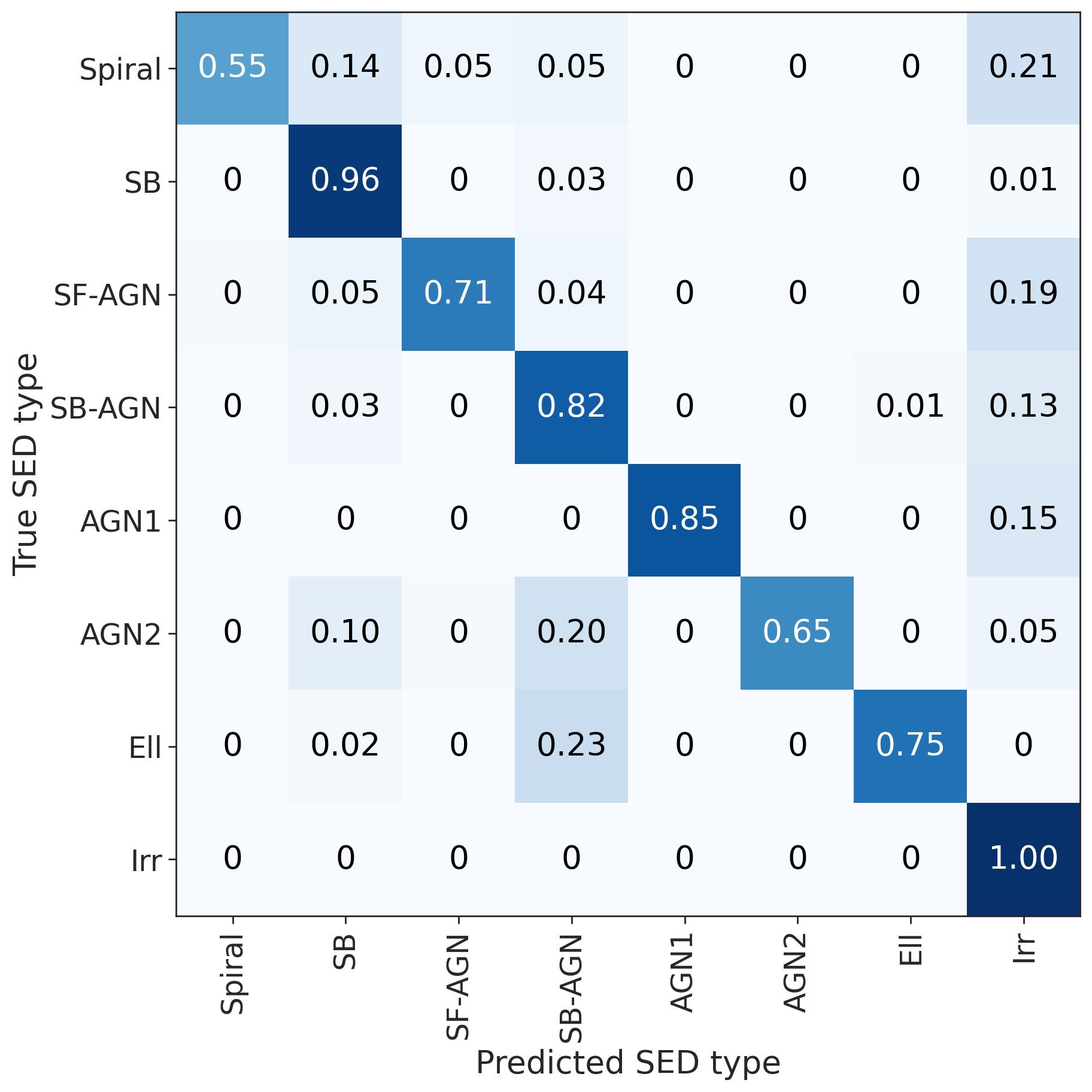}
    \caption{Test set confusion matrix for SED type classification. Each row of the matrix represents the fraction of samples in an actual SED type, while each column represents the fraction of samples in a predicted SED type. The superimposed numbers indicate the fraction of objects of a class in that particular position. For example, considering the second row (SB): $96\%$ of SBs are correctly classified as such, $3\%$ as SB-AGN and $1\%$ as Irregular.}
    \label{fig:CM}
\end{figure}

Since the {\tt XGBoost} input data and the train-test split used are the same as in the photometric redshift estimation, we used the same hyperparameters (Table \ref{table:hyperparam}).
The accuracy obtained on the test set is $96.8\%$, meaning that $3\,529\,289$ galaxies out of $3\,645\,481$ are correctly classified. Some significant misclassifications (see the confusion matrix in Fig. \ref{fig:CM}) are explained by considering the nature of the SEDs and the photometry available. For example, it is reasonable that AGN2 and SB-AGN, both obscured in the optical, are misclassified as SB, since it would be necessary to have spectroscopic data or photometric observations in the mid-IR (rest-frame) and X-ray, to identify their AGN nature.

\begin{figure}
    \centering
    \includegraphics[width=\linewidth]{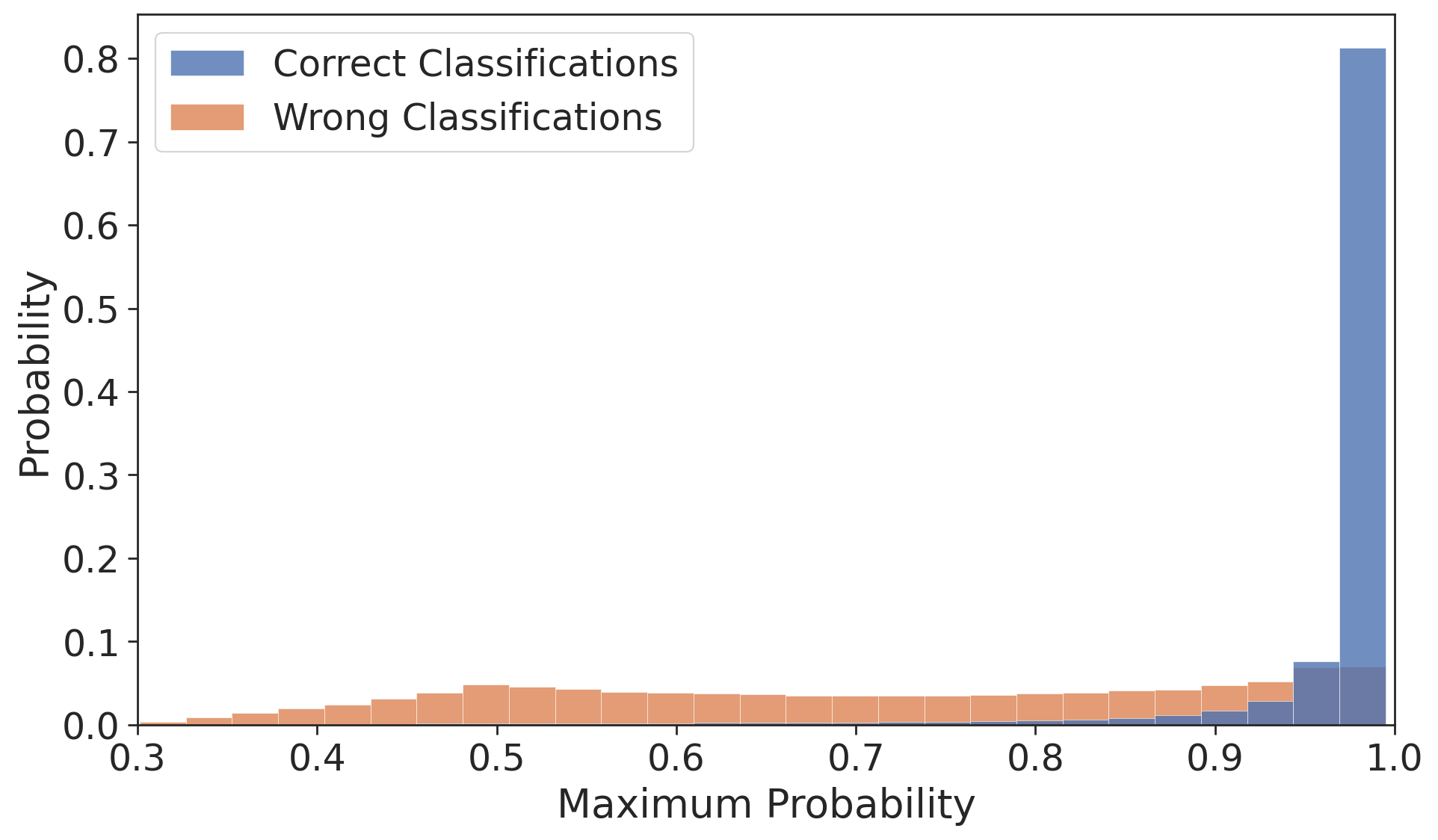}
    \caption{Maximum probability distribution for correct (blue) and wrong (orange) classifications. The sum of the heights of each histogram is one.}
    \label{fig:prob_dist}
\end{figure}

The distribution of the maximum probability per object for correct and wrong classifications is shown in Fig. \ref{fig:prob_dist}. While for the mistakenly classified samples it is a rather flat distribution, for correct predictions it is strongly peaked at values of maximum probability larger than $\approx0.95$. This means it is possible to improve the spectral type prediction accuracy even more, by simply considering objects with large maximum probability, while losing a negligible number of correct predictions. For example, by keeping only objects with maximum probability $>0.8$, the number of misclassified objects goes down by $61\%$, while dropping only $3.7\%$ of the correctly classified ones. The resulting accuracy is $98.7\%$.
\subsection{Identifying HIEROs}\label{sec:identifyHIEROs}
Having settled the general framework for the estimations of photometric redshifts and spectral types, in this section we present the results of the approach applied to the most relevant population in our study, i.e. the red and massive galaxies at distances larger than $z \approx 3$. We are selecting them as HIEROs (Eq. \ref{eq:hieros} and Fig. \ref{fig:wang_cat}).

As these massive galaxies are challenging to spectroscopically confirm, to obtain a more realistic estimate of the identification performance of this population, we train the gradient-boosting algorithm with a more realistic spectroscopic completeness: we utilized the training set as described in the previous sections but retaining only 500 HIEROs, which accounts for approximately 1/46 of the total number of HIEROs that would be available.
We consider this a conservative approach as:
\begin{itemize}
    \item we assume that a robust sample of HIERO that we could use as training set will be available in the COSMOS field, thanks to the very precise and accurate photometric redshifts derived from the already available and future multi-wavelength observations (\citetalias{Weaver2022}; \citealt{Casey2022});
    \item we assume that in the next years we will have access to spectroscopic redshifts for HIEROs thanks to JWST, as there are already photometric candidates suitable for follow-up \citep[i.e., 138 objects similar to HIEROs in an area of $38.8\,{\rm arcmin}^2$][]{perezgonzalez}.
\end{itemize}

As there is clearly some stochasticity in the sampling of the HIEROs in the training set, the results reported are the average across multiple runs.

In Table \ref{table:reg_hieros} their identification performance is shown in terms of photo-$z$ and SED type classification. 

\begin{figure*}
    \centering
    \includegraphics[width=\linewidth]{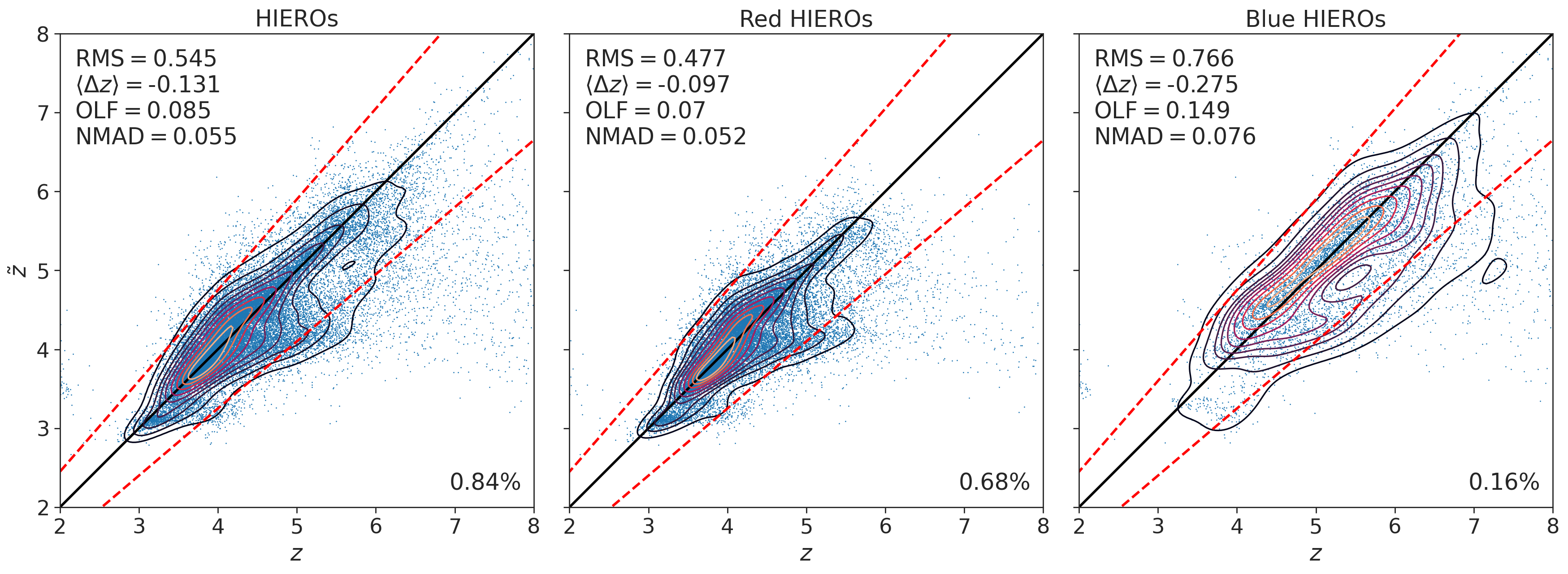}
    \caption{Contour plot of {\tt XGBoost} predictions ($\Tilde{z}$) versus catalog redshifts ($z$) colored by density in the $(z, \Tilde{z})$ space for HIEROs (left panel), red HIEROs (middle panel) and blue HIEROs (right panel), all having $\HE\leq26$ ; lighter contour colors mean higher density. The red dashed lines show $\Tilde{z} = z\pm  0.15(1 + z)$. The percentage of objects with respect to $N_{\rm test}$ is indicated in the bottom right corner of each panel.}
    \label{fig:res_hieros}
\end{figure*}

\begin{figure}
    \centering
    \includegraphics[width=\linewidth]{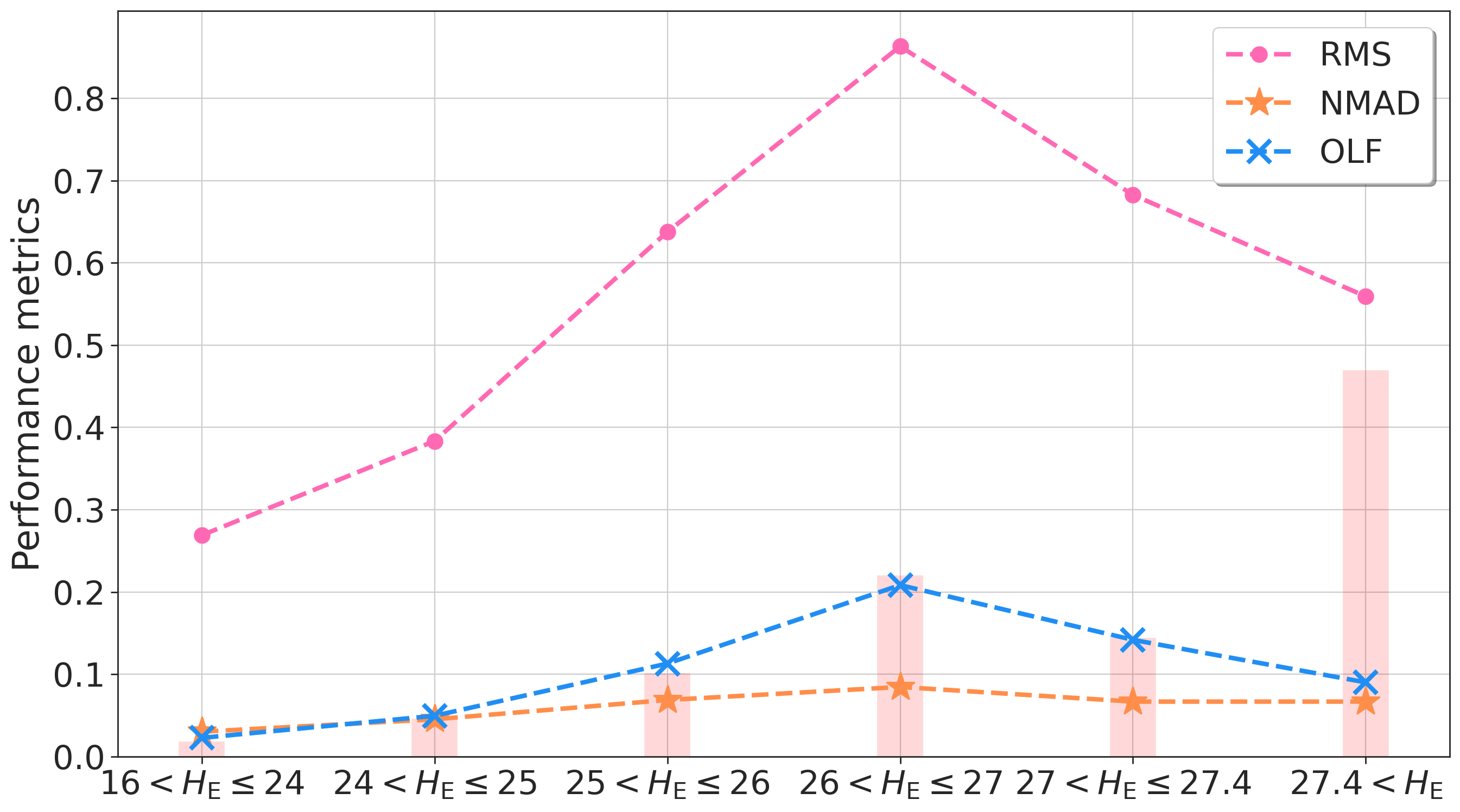}
    \caption{$\rm RMS$, $\rm OLF$ and $\rm NMAD$ values for HIEROs of different $\HE$-band magnitudes. The vertical pink bands indicate the fraction of objects belonging to the group with respect to the total number of HIEROs.}
    \label{fig:hieros_perf}
\end{figure}

The regression performance is compared to that reported in \citet{Wang2016}, where the $\rm NMAD$ is approximately $0.1$.
Even for this population, the performance obtained in this work (HIEROs $\rm NMAD=0.068$) is encouraging, especially for the brightest galaxies ($\HE\leq26$), as shown in Figs. \ref{fig:res_hieros} and \ref{fig:hieros_perf}.
Clearly, the advantage of the Euclid Deep Fields survey will be the HIEROs' much larger statistical relevance compared to previous surveys.
\begin{table*}[t!]
\caption{Redshift and Spectral type prediction performance for HIEROs. $N$ indicates the number of objects in the selection. ‘‘Other’’ indicates the classes of objects with $\HE-[4.5]<2.25$.}
\label{table:reg_hieros}
\centering          
\begin{tabular}{c c c c c c c}
\hline
\multirow{2}{*}{}&\multirow{2}{*}{$N$}&\multicolumn{4}{c}{Photo-$z$}&SED classification\\
       &&RMS  &$\langle \Delta z\rangle$& OLF &NMAD& Accuracy \\
\hline
\hline
HIEROs&$184\,441$&0.654&$-$0.084&0.123&0.068&86.20\% \\
Red HIEROs&$51\,092$&0.668&$-$0.059&0.143&0.073&86.28\%  \\
Blue HIEROs&$89\,996$&0.690&$-$0.119&0.126&0.061&86.28\% \\
\vspace{2mm}
Other&$3\,461\,040$&0.341&$-$0.013&0.057&0.038&97.20\%\\
$\HE\leq26$ HIEROs&$30\,632$&0.545&$-$0.131&0.085&0.055&92.68\% \\
$\HE\leq26$ Red HIEROs&$24\,645$&0.477&$-$0.097&0.07&0.052&94.44\% \\
$\HE\leq26$ Blue HIEROs&$5987$&0.766&$-$0.275&0.149&0.076&85.44\% \\
\hline
\end{tabular}
\end{table*}

We also report in Table \ref{table:CompPurity} the completeness and fraction of contaminants in different redshift ranges. The redshift range where we expect to find the vast majority of these objects, i.e. $3\leq z<7$, is complete at $99.4\%$, with a contaminant fraction of only $5\%$, mostly coming from higher-redshift galaxies.

\begin{figure*}
    \centering
    \includegraphics[width=\linewidth]{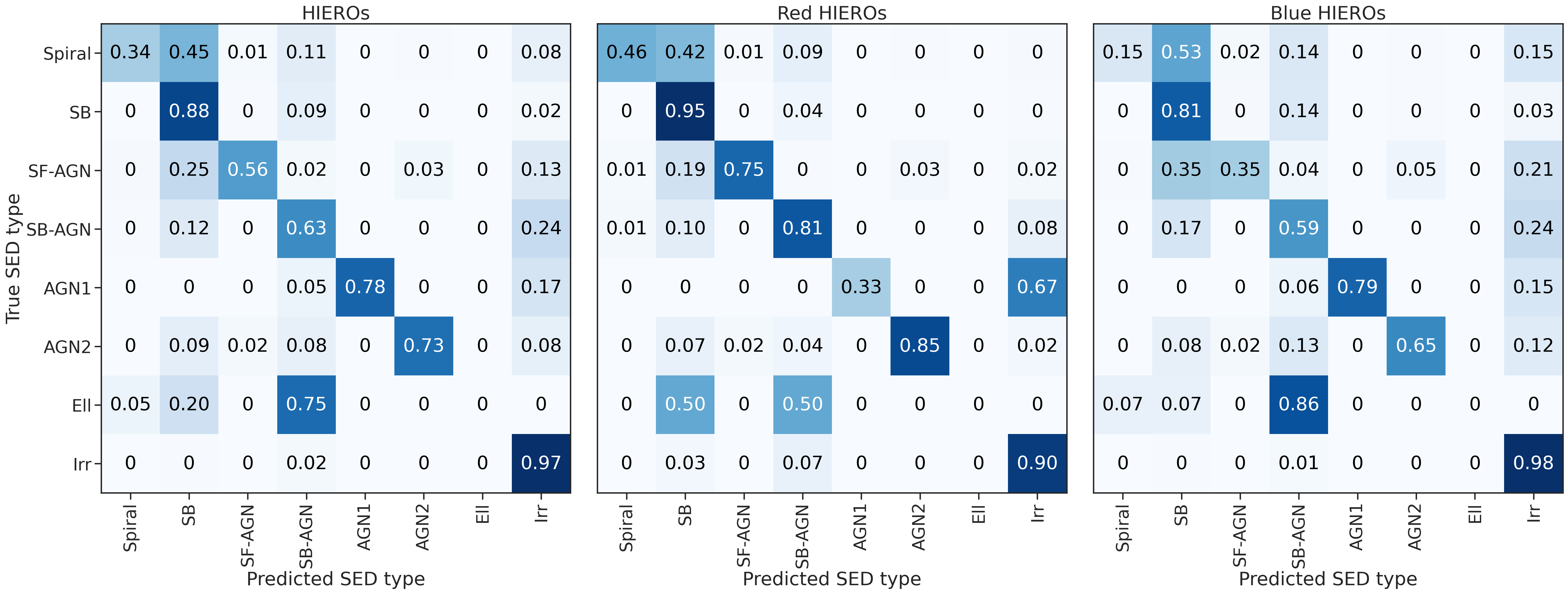}
    \caption{SED type classification confusion matrix for HIEROs, red HIEROs and blue HIEROs. The high confusion for ellipticals is explained by noting that there are only 20 of them in the HIERO selection (six in the red HIEROs and 14 in the blue HIEROs), none of them correctly identified.}
    \label{fig:cm_hieros}
\end{figure*}

Lastly, the spectral type classification, although less accurate than for non-HIEROs ($86.20\%$ vs. $97.20\%$), shows (Fig. \ref{fig:cm_hieros}) a similar confusion between classes to the classifications for all the objects in the catalog (Fig. \ref{fig:CM}).
\setlength{\tabcolsep}{4pt}
\begin{table}
\caption{Confusion matrix for photometric redshift ranges for HIEROs. Here $z$ and $\tilde{z}$ indicate true and predicted redshifts, respectively. The diagonal entries represent the completeness of the samples, while the off-diagonal terms their contaminants. The sum along rows of the off-diagonal terms equals $1-purity$. In parenthesis we report the numbers of objects.}      
\label{table:CompPurity}    
\centering                                      
\begin{tabular}{c | c c c}      
\hline              
 &$z<3$&$3\leq z<7$&$7\leq z$\\
\hline                           
$\tilde{z}<3$&37.2\% (1537)&22.1\% (438) &0\% (9)\\
$3\leq \tilde{z}<7$&1.4\% (2590)&99.4\% ($169\,707$)&3.6\% (6442)\\
$7\leq \tilde{z}$&0\% (2)&15.2\% (566)&32.8\% (3150)\\
\hline                                             
\end{tabular}
\end{table}

%% file: Sections/discussion.tex
\section{Summary and discussion}\label{sec:discussion}
\subsection{A new approach to identifying the most elusive and massive systems with \Euclid at high-$z$}
In this work we have attempted a photometric approach to predict whether \Euclid will provide enough information to answer one of the key scientific questions in galaxy formation and evolution studies: what is the role of distant obscured galaxies in the build up of today’s large scale structures?

Such sources have proven to be elusive in optical surveys, and current near-IR surveys are deep enough to detect them only in very small sky areas. Since these optically faint objects are rare, large area surveys to sufficient depths are necessary to provide a statistical census of such a population. \Euclid will potentially provide all the ingredients to recover these missing galaxies at high redshift (e.g., $z>3$). However, it has been unclear if the photometric information in hand will be sufficient to identify and characterize them, given the lack of bright spectroscopic features observable by \Euclid at these distances (i.e. $\lambda_{\rm rest-frame}<0.5\, \micron$).

Our goal was therefore to assess the capability of \Euclid + ancillary data to identify these distant obscured objects.

We carried out the following analysis:
\begin{itemize}
    \item We adopted the Euclid Deep Fields simulated catalog from the \spr simulation as a basis for our analysis and showed that it includes a massive and dusty population, as selected by the criterion $\HE - [4.5] > 2.25$. $98\%$ of these simulated objects (the HIEROs) are indeed at  $3<z<7$ and contribute significantly ($\approx 93\%$) to the massive-end of the stellar mass function at $z>4$.
    \item We implemented a general fast and accurate machine learning technique optimized for $z\ge2$ galaxy identification, based on photometric data from \Euclid, Rubin and \textit{Spitzer}, finding that only a $\approx 10\%$ subset of the total observed galaxies with spectroscopic redshifts (or reliable photometric redshifts) is required in order to obtain good performance ($\rm RMS=0.363$, $\rm OLF=0.061$, $\langle\Delta z\rangle=-0.017$, $\rm NMAD=0.039$ over $2\leq z<8$).
    We observed that this photo-$z$ precision is comparable to that of the COSMOS2020 catalog, obtained with state-of-the-art traditional SED-fitting methods with almost four times the number photometric bands.
    \item We applied the same methods for spectral type classification, distinguishing between eight different spectral types (i.e. spiral, starburst, star-forming AGN, starburst AGN, type-1 AGN, type-2 AGN, elliptical and irregular), obtaining an accuracy of $96.8\%$.
    \item We evaluated the identification performance for objects within the HIEROs selection and found a photo-$z$ $\rm OLF=0.123$ and $\rm NMAD=0.068$; the SED classification accuracy is $86.2\%$. 
    Evaluating only the brightest ($\HE<26$) HIEROs, we obtained an $\rm OLF=0.085$, an ${\rm NMAD}=0.055$ and a $92.68\%$ SED classification accuracy.
    \item We determined the completeness and the fraction of contaminants for the HIEROs photo-$z$. In the range $3\leq z<7$, we estimated a completeness of $99.4\%$ and a contaminant fraction of $5\%$, the majority of which are placed at $z>7$.
\end{itemize}
All these results suggest that by leveraging \Euclid, Rubin and \textit{Spitzer} photometric data, and by taking the approach described in this paper, it will be possible to constrain the contribution of the dusty and massive galaxies at $ z \approx3-7$ to the mass functions and, hopefully, to the star-formation rate density.

\subsection{Future perspectives}
Under suitable conditions, gradient-boosting (like many other supervised machine learning techniques) is a very competitive tool for photometric redshift and spectral type estimation. However, its successful application depends on the availability of a large enough training set that is representative of the populations under consideration. Its effectiveness hence lies particularly within large photometric surveys, some of which include spectroscopic data for subsets of the photometric catalogs.
One considerable problem for these methods is the difficulty in extrapolating to regions of the features space not properly represented in the training data; the distribution of magnitudes and colors of the training set has to be as close as possible to the ones in the target set.

Lastly, the good performance demonstrated here relies heavily on the representativeness of the \spr simulation with respect to the \Euclid observations. 
It has been shown in several papers (\citealt{SPRITZ} and \citeyear{Bisigello2022}) that \spr is in agreement with a large set of observations at different wavelengths. However, the simulation as also some limitation. For example, the number of SED templates included in the simulation are limited and results at number densities $z>3$ are mainly based on extrapolation, even if they have been largely tested. Therefore, the results shown here may be prone to biases or slightly optimistic compared to those that will be obtained once real data become available. 

To try to mitigate this limitation and have a more robust forecast, future work could then apply the methods of this paper on different simulated data sets, for example {\tt MAMBo} (Mocks with Abundance Matching in Bologna, \citealt{MAMBO?}) and compare the results obtained with those of the official \Euclid pipelines.

Furthermore, some additional strategies can also be taken in order to improve the predictions. Among these, the gradient-boosting algorithm is easily extendable in order to provide a probability distribution even on redshifts.

After implementing the most effective photometric pipeline to real data, a promising selection of candidate galaxies will be obtained. To confirm their nature, spectroscopic follow-up will be necessary, for example from the Extremely Large Telescope, ALMA or the James Webb Space Telescope.

%% file: Sections/appendix.tex
\begin{appendix}
\section{Missing detections handling}
We show some clarifying examples of the handling of missing detections used in this work in Table \ref{table:NaN_Handling}.
\begin{table}[H]
\caption{Three examples of data points (rows) with non-detections to clarify the missing numbers handling adopted in this work. 
The upper table indicates the original values of magnitudes, colors and ratios, while the lower table give the corresponding values after the missing detection handling.
$M_1$ and $M_2$ indicate two different bands magnitudes, while $M_1^{(2\sigma)}$ and $M_2^{(2\sigma)}$ are their $2\sigma$ detection limits, as reported in Table \ref{table:depths}.
}              
\label{table:NaN_Handling}  
\begin{tabular}{c c c c}
\hline
$M_1$ & $M_2$ & $M_1-M_2$ & $M_1/M_2$ \\
\hline\hline
NaN& 27.17&NaN&NaN\\
25.72& NaN&NaN&NaN\\
NaN&NaN&NaN&NaN\\
\hline
\multicolumn{4}{c}{$\big\Downarrow$}\\
\hline
$M_1$ & $M_2$ & $M_1-M_2$ & $M_1/M_2$ \\
\hline\hline
$M_1^{(2\sigma)}$& 27.17&$M_1^{(2\sigma)}$&$M_1^{(2\sigma)}$\\
25.72& $M_2^{(2\sigma)}$& $-M_2^{(2\sigma)}$&1/$M_2^{(2\sigma)}$\\
$M_1^{(2\sigma)}$&$M_2^{(2\sigma)}$&NaN&NaN\\
\hline
\end{tabular}
\end{table}
\section{Photo-$z$ performance}
Here we report the detailed {\tt XGBoost} results for photometric redshifts (evaluated on the test set), for different redshift bins, SED types, $i$-band AB magnitudes and number of missing detections (shown graphically in Fig. \ref{fig:total_evals}).
\begin{table}
\caption{{\tt XGBoost} Test Set RMS, bias, OLF and NMAD for different redshift intervals, SED types, number of missing bands and $i$-band magnitudes. The redshift range $2 \leq z < 8$ comprises all the galaxies in the test set. Objects with $i > 29.3$ are not detected in the $i$-band.}              
\label{table:1090res}             
\centering
\begin{tabular}{c c c c c c}
\hline
$z$& $N$&RMS  & OLF &NMAD&$\langle \Delta z \rangle$\\
\hline
\hline
$2\leq z<8$     &  3645481 &      0.363 &     0.060 &      0.039 & $-$0.017 \\ 
$2\leq z<3$     &  2047751 &      0.254 &     0.047 &      0.037 &  0.057 \\ 
$3\leq z<4$     &   860381 &      0.365 &     0.078 &      0.046 & $-$0.049 \\ 
$4\leq z<5$     &   407092 &      0.368 &     0.049 &      0.040 & $-$0.067 \\ 
$5\leq z<6$     &   195186 &      0.532 &     0.089 &      0.044 & $-$0.230 \\ 
$6\leq z<7$     &    91505 &      0.749 &     0.124 &      0.037 & $-$0.372 \\ 
$7\leq z<8$     &    43566 &      1.289 &     0.207 &      0.044 & $-$0.718 \\ 
\hline
\hline
SED Type &$N$&RMS  & OLF &NMAD&$\langle \Delta z \rangle$\\
\hline
\hline
Spiral          &    24719 &      0.461 &     0.057 &      0.030 & $-$0.045 \\ 
SB              &   293497 &      0.665 &     0.144 &      0.044 &  0.007 \\ 
SF-AGN          &    42778 &      0.567 &     0.085 &      0.026 & $-$0.109 \\ 
SB-AGN          &   243550 &      0.430 &     0.051 &      0.030 & $-$0.044 \\ 
AGN1            &    72714 &      0.558 &     0.137 &      0.044 & $-$0.062 \\ 
AGN2            &    23009 &      0.480 &     0.102 &      0.041 & $-$0.006 \\ 
Ell             &    21535 &      0.219 &     0.027 &      0.038 &  0.052 \\ 
Irr             &  2923679 &      0.297 &     0.050 &      0.040 & $-$0.015 \\ 
\hline
\hline
$i$-band &$N$&RMS  & OLF &NMAD&$\langle \Delta z \rangle$\\
\hline
\hline
$16<i\leq 25$   &   209415 &      0.045 &     0.000 &      0.008 & $-$0.010 \\ 
$25<i\leq 26$   &   352218 &      0.086 &     0.001 &      0.013 & $-$0.013 \\ 
$26<i\leq 27$   &   683341 &      0.185 &     0.008 &      0.023 & $-$0.018 \\ $27<i\leq 28$   &  1245136 &      0.332 &     0.043 &      0.050 & $-$0.016 \\ 
$28<i\leq 29.3$ &  1080295 &      0.516 &     0.141 &      0.088 & $-$0.019 \\ 
$29.3<i$        &    75076 &      0.631 &     0.119 &      0.072 & $-$0.040 \\ 
\hline
\hline
Missing Bands &$N$&RMS & OLF &NMAD&$\langle \Delta z \rangle$ \\
\hline
\hline
0 &   741823 &      0.082 &     0.001 &      0.014 & $-$0.009 \\ 
1  &   930708 &      0.201 &     0.011 &      0.027 & $-$0.018 \\ 
2  &   608234 &      0.391 &     0.059 &      0.051 & $-$0.020 \\ 
3  &   420111 &      0.436 &     0.082 &      0.065 & $-$0.020 \\ 
4  &   389306 &      0.431 &     0.095 &      0.079 & $-$0.016 \\ 
5  &   271637 &      0.512 &     0.156 &      0.096 & $-$0.018 \\ 
6  &   165679 &      0.632 &     0.209 &      0.109 & $-$0.019 \\ 
7  &   117983 &      0.688 &     0.215 &      0.112 & $-$0.032 \\
\hline
\end{tabular}
\end{table}
\section{Photo-$z$ feature importance}\label{app:fi}
\begin{figure}
    \centering
    \includegraphics[width=\linewidth]{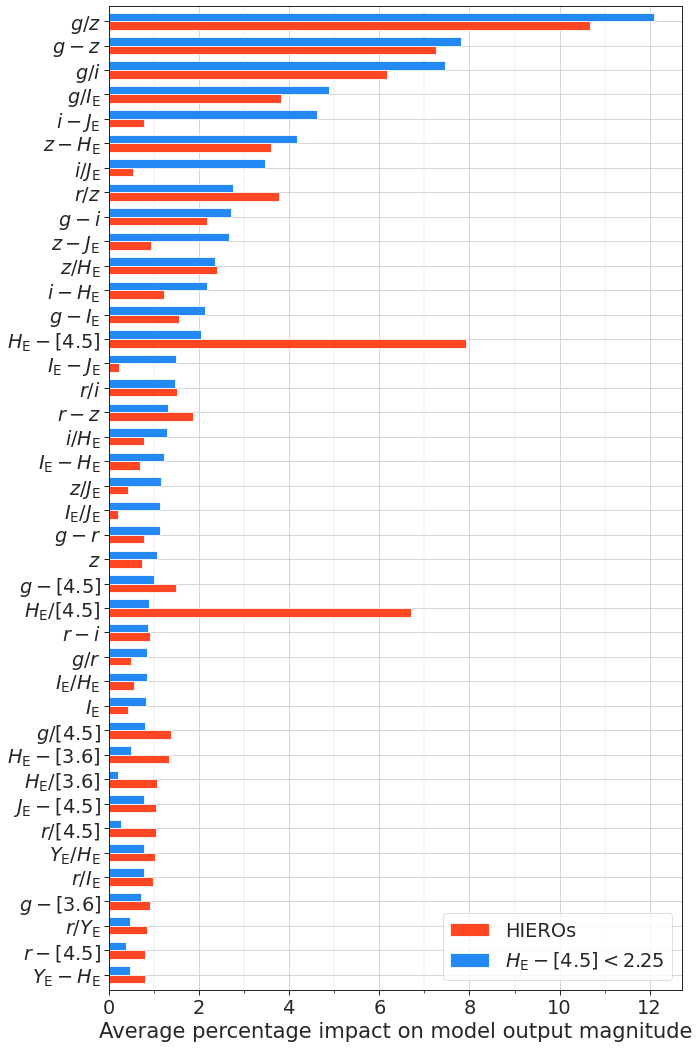}
    \caption{Feature rankings from the {\tt XGBoost} model for HIEROs (red) and other objects (blue). We used the average of the absolute value of the normalized (by predicted redshift) SHAP values. This corresponds to the average percentage impact on model outputs. For example, on average, $8\%$ of the model output for HIEROs is based on the $\HE-[4.5]$ color, while for other objects the impact of this feature is only around $2\%$.}
    \label{fig:shapley_bar}
\end{figure}

Figure \ref{fig:shapley_bar} shows the most important features for redshift estimation (evaluated on the test set) for HIEROs and other objects, in terms of Shapley values (SHAP, \citealt{SHAP}). These can be both positive and negative, and indicate the contribution of every feature to the model output, as for every data point the sum of its Shapley values is the predicted redshift. Here the sum of the modulus of these values is normalized to one.

While the redshift prediction is mostly based on colors and magnitude ratios, the feature importance order of the two populations is quite different, with the HIEROs redshift estimates being more strongly reliant on the near-infrared magnitudes. The colors and ratios of the same magnitude pairs are ranked closely.
Note, however, the number of missing detections per band (Table \ref{table:detections_per_band}) and the consequent reduced variability of the magnitudes, which clearly influences the importance.
\section{Contaminants analysis}\label{sec:contaminants}
Throughout the paper, we considered only objects at $z>2$, under the assumption that the selection of high-redshift objects from official photo-$z$ pipelines would yield a clean data set.
In this appendix we analyze potential contamination issues from low-redshift ($z < 2$) sources, as well as brown dwarfs, in the selection of $3<z<7$ HIEROs candidates. 

To address the first concern, we apply a dedicated binary classifier using the same XGBoost algorithm and input features as described in the main body of the paper (see Sect. \ref{sect:SED_class}), with the goal of differentiating between objects at $z<2$ and $z>2$. 
This classifier is trained on a data set comprehending objects down to $z=0$ and $500$ HIEROs (see Sect. \ref{sec:identifyHIEROs}).
This training set consists of $2\,815\,072$ galaxies, the vast majority of which are at $z<2$ (see, for example, Fig. \ref{fig:Catalogzmass}, top panel).

The objective is to estimate the degree of contamination from low-$z$ sources in the HIEROs selection (Eq. \ref{eq:hieros}), essentially quantifying the number of galaxies with $z < 2$ that both fit within the color-color selection and that are predicted by the classifier at $z > 2$.
We report the completeness and fraction of contaminants obtained from this analysis in Table \ref{table:low-z_contaminants}, reminding the reader that in the testing phase we consider only observations with at least four bands with ${\rm S/N} > 2$ and at least one band with ${\rm S/N} > 5$. 
\setlength{\tabcolsep}{4pt}
\begin{table}
\caption{Confusion matrix for photometric redshift ranges for HIEROs. Here $z$ and $\tilde{z}$ indicate true and predicted redshift class, respectively. The diagonal entries represent the completeness of the samples, while the off-diagonal terms their contaminants. In parenthesis we report the numbers of objects.}   
\label{table:low-z_contaminants}    
\centering                                      
\begin{tabular}{c | c c }      
\hline              
 &$z<2$&$z\ge 2$\\
\hline                           
$\tilde{z}<2$&59.7\% (2882)&39.9\% (1914) \\
$\tilde{z}\ge 2$&1.1\% (1948)&98.9\% ($182\,527$)\\
\hline
\end{tabular}
\end{table}

These results underscore the classifier's effectiveness in distinguishing between $\HE-[4.5]> 2.25$ sources with redshifts above and below $2$, showcasing high completeness at $z>2$ ($98.9\%$) and high purity, thereby minimizing the risk of contamination (kept at only $1.1\%$). 

To assess the contamination by brown dwarfs, we simulate \Euclid magnitudes based on the models of L and T dwarf from \citet{Burrows2006}. These templates have effective temperatures ranging from 700 K to 2300 K, metallicities from [Fe/H]=$-0.5$ to 0.5, and gravities fom $10^{4.5}$ to $10^{5.5}\,\rm cm\, s^{-2}$. We do not simulate their spatial distribution, as this goes beyond the scope of this paper, but we derive magnitudes assuming they are located from 1 to 288 pc away from us, with steps of 1 pc. At larger distances, all brown dwarfs models correspond to ${\rm S/N} < 3$ in all \Euclid filter. We kept all mocks that results in a ${\rm S/N} > 3$ in at least one \Euclid filter, for a total of $33\,866$ simulated brown dwarfs. We then scatter their magnitudes using the same procedure reported in Sec. \ref{sec:photerr}.
We find that less than $1\%$ of these brown dwarfs fall into the selection $\HE-[4.5]> 2.25$ and thus the contamination from these sources in the HIEROs selection (applied to all objects in our catalog with $z>2$) is below $0.01\%$.

Consequently, we argue that the conclusions drawn in the main body of the paper remain largely unaffected even with the possible inclusion of $z < 2$ galaxies and brown dwarfs in the analysis.
\section{Identifying HIEROs with shallower Rubin depths}
The analysis presented in this work assumes Rubin depths after 10 years of observations. To evaluate the efficacy of our search for the reddest high-redshift galaxies before that time, we run the same tests described above with shallower depths (Table \nolinebreak\ref{table:depths_years}), corresponding to one and four years of Rubin observations (\citealt{RubinDepths}). In all our runs, we considered 500 HIEROs in the training set.

As expected, the photo-$z$ performance improves with each passing year, while also benefiting from a larger sample of HIEROs in the test set. The decrease in RMS and $\langle \Delta z \rangle$ from year one to year 10 is a clear indication that the accuracy of the predictions has improved significantly over time. However, the most notable improvement is in the catastrophic outlier fraction metric, which decreased from $0.254$ in year one to $0.141$ in year 10. This is a significant improvement, and it suggests that deeper observations are particularly important in reducing exceptionally wrong predictions.
\begin{table}
    \caption{Expected $5\sigma$ Rubin depths after one and four years of observations. To be compared with the ones after 10 years, reported in Table \nolinebreak\ref{table:depths}.}
\label{table:depths_years}     
\centering                            
\begin{tabular}{c c c c c c}          
\hline           
 Rubin year& Rubin/\textit{u}& Rubin/\textit{g}
 &Rubin/\textit{r}&
 Rubin/\textit{i} &
 Rubin/\textit{z}\\ 
\hline  \hline 
1 &24.5 & 25.7 & 25.8 & 25.1 &24.4\\
4 &25.2 & 26.4 & 26.5 & 25.8 &25.1\\
\hline                   
\end{tabular}
\end{table}
\begin{table}
\caption{Redshift prediction performance for HIEROs, as obtained considering Rubin depths after one and four years of observations. $N$ indicates the number of objects in the selection. For ease of comparison, we report again the results after 10 years of Rubin observations, already presented in Table \ref{table:reg_hieros}.}
\label{table:reg_hieros_years}
\centering          
\begin{tabular}{c c c c c c}
\hline
Rubin year & $N$&RMS  & OLF &NMAD&$\langle \Delta z \rangle$\\
\hline
\hline
1 & $115\,303$ & 0.864 &0.301&0.112&$-$0.064\\
4 & $115\, 371$ & 0.858&0.229&0.110&$-$0.031\\
10 & $184\,441$ & 0.654&0.123&0.068&$-$0.084\\
\hline
\end{tabular}
\end{table}
\section{Rubin/$y$}
We did not include in our analysis the Rubin/$y$ band ($2\sigma$ depth of $27.2\,\mathrm{mag}$), as it was determined to have negligible impact on our results. We address this point by showing in Table \ref{table:with_y} the improvement in performance between the data set with and without the Rubin/$y$ band.
This results demonstrate that the inclusion of this band yields little improvement on our results across most redshifts and magnitude intervals.
\begin{table}[h!]
\caption{{\tt XGBoost} Improvement in test set RMS, OLF and NMAD, bias for different redshift intervals $i$-band magnitudes, as obtained with the inclusion of the Rubin/$y$ band. Objects with $i > 29.3$ are not detected in the $i$-band.}
\label{table:with_y}
\centering          
\begin{tabular}{c c c c c c}
\hline
$z$&RMS & OLF &NMAD&$\langle \Delta z \rangle$\\

\hline
\hline
$2\leq z<8$ &      0.011 &     0.005 &     0 & $-$0.003 \\
$2\leq z<3$ &     0 &     0.001 &      0.001 &  0.001 \\
$3\leq z<4$ &      0.010 &     0.006 &      0.001 & 0 \\
$4\leq z<5$ &      0.007 &     0.004 &     0 & 0\\
$5\leq z<6$ &      0.026 &     0.012 &      0.001 & $-$0.028\\
$6\leq z<7$ &0.069 &     0.029 &      0.001 & $-$0.051 \\
$7\leq z<8$ &      0.140 &     0.037 &      0.004 & $-$0.106 \\
\hline
\hline
$i$-band &RMS  & OLF &NMAD&$\langle \Delta z \rangle$\\
\hline
\hline
$16<i\leq 25$ &     $-$0.002 &    $-$0.001 &     0 & $-$0.001 \\
$25<i\leq 26$ &     $-$0.001 &    0 &     0 & 0 \\
$26<i\leq 27$ &      0.015 &     0.002 &     0 & $-$0.001 \\
$27<i\leq 28$ &      0.019 &     0.005 &      0.001 & $-$0.002 \\
$28<i\leq 29.3$ &      0.020 &     0.014 &      0.004 & $-$0.003 \\
$29.3<i$ &     $-$0.008 &    $-$0.009 &      0.005 & $-$0.014 \\
\hline
\end{tabular}
\end{table}
\end{appendix}